\documentclass[aoas]{imsart}

\RequirePackage{amsthm,amsmath,amsfonts,amssymb}
\RequirePackage[authoryear]{natbib}
\usepackage{algorithm, graphicx}
\RequirePackage[colorlinks,citecolor=blue,urlcolor=blue]{hyperref}
\usepackage{ulem}

\def\argmax{\arg\max}

\newcommand{\wh}{\widehat}

\startlocaldefs
\numberwithin{equation}{section}

\endlocaldefs

\begin{document}

\begin{frontmatter}
\title{Causal Discovery on Dependent Mixed Data with Applications to Gene Regulatory Network Inference}
\runtitle{Causal Discovery on Dependent Mixed Data}

\begin{aug}
\author[A]{\fnms{Alex}~\snm{Chen}}
\and
\author[A]{\fnms{Qing}~\snm{Zhou}\ead[label=e2]{zhou@stat.ucla.edu}}
\address[A]{Department of Statistics and Data Science,
University of California, Los Angeles  \printead[presep={\\ }]{e2}}
\end{aug}

\begin{abstract}
Causal discovery aims to infer causal relationships among variables from observational data, typically represented by a directed acyclic graph (DAG). Most existing methods assume independent and identically distributed observations, an assumption often violated in modern applications. In addition, many datasets contain a mixture of continuous and discrete variables, which further complicates causal modeling when dependence across samples is present. 
To address these challenges, we propose a de-correlation framework for causal discovery from dependent mixed data. Our approach formulates a structural equation model with latent variables that accommodates both continuous and discrete variables while allowing correlated Gaussian errors across units. We estimate the dependence structure among samples via a pairwise maximum likelihood estimator for the covariance matrix and develop an EM algorithm to impute latent variables underlying discrete observations. A de-correlation transformation of the recovered latent data enables the use of standard causal discovery algorithms to learn the underlying causal graph.
Simulation studies demonstrate that the proposed method substantially improves causal graph recovery compared with applying standard methods directly to the original dependent data. We apply our approach to single-cell RNA sequencing data to infer gene regulatory networks governing embryonic stem cell differentiation. The inferred regulatory networks show significantly improved predictive likelihood on test data, and many high-confidence edges are supported by known regulatory interactions reported in the literature.

\end{abstract}


\end{frontmatter}

\section{Introduction}
Causal discovery aims to identify the underlying causality among variables from observed data. It is usually formalized through the framework of directed acyclic graphs (DAGs), where the nodes of a DAG represent random variables and edges encode causal relationships between variables. Specifically, let $V = \{ V_1,\ldots,V_p\}$ denote a set of $p$ random variables, and let $\mathcal{G} = (V,E)$ be a DAG with variable (node) set $V$ and edge set $E$ such that an edge $i \rightarrow j \in E$ implies $V_i$ is a direct cause of $V_j$. A causal discovery method learns the structure (i.e., the edge set) of the underlying DAG that best explains the observed conditional independencies among the variables. The directed edges in the learned graph indicate which variables directly influence others, thereby distinguishing association from causation.

Causal discovery methods are commonly separated into three categories: constraint-based, score-based, and hybrid approaches. \citet{glymour2019review} and \citet{nogueira2022methods} provide comprehensive reviews of causal discovery methods. Here, we briefly review the most classical algorithms from each category. Constraint-based methods perform a sequence of conditional independence tests to systematically identify and orient edges in a graph. Typically, these methods begin with a complete undirected graph over the observed variables and then iteratively prune edges when two variables are conditionally independent given a set of other observed variables. The result of the pruning stage is a skeleton, which is then oriented using logical orientation rules derived from graph properties such as v-structures and acyclicity as introduced by \citet{10.5555/2074158.2074204}. The most well-known constraint-based method is the PC algorithm \citep{spirtes2000causation}. Score-based methods search over a space of graphs to identify a DAG or an equivalence class of DAGs that maximizes a predefined scoring function. A commonly used example is the Bayesian Information Criterion (BIC) \citep{BIC}, which balances goodness of fit with model complexity by penalizing overly complex graphs. Common examples of score-based methods include Greedy Equivalence Search (GES) \citep{chickering2002optimal} and Greedy Hill Climbing \citep{gamez2011learning}, both of which iteratively modify the graph in order to improve the score. Hybrid methods integrate both approaches, constraint and score-based learning, in order to leverage different strengths of each approach. One example is the Max-min Hill-Climbing (MMHC) algorithm \citep{tsamardinos2006max}. MMHC first uses conditional independence tests to identify a skeleton. Then, a score-based search is used to orient the edges in the learned skeleton by selecting directions that maximize the scoring criterion.

\subsection{Motivation}
Most causal discovery methods rely on a key assumption: observations (units) are independent and identically distributed (i.i.d.). However, this assumption is often violated in real-world applications, where dependence among samples naturally arises. For instance, in social network analysis, individuals' behaviors are influenced by their social connections, such as friends, families, or colleagues, resulting in interdependent features across individuals. Another layer of complexity arises when the data consists of both continuous and discrete variables. Social network data can contain a mixture of continuous engagement metrics, binary indicators, and categorical demographic variables. In genomics, some variables are typically represented as discrete states, such as the presence or absence of a protein in the cell, while other variables, say the expression levels of genes, vary continuously. It is also a common practice to discretize gene expression levels depending on the data quality and generation mechanisms. Mixed data can be particularly challenging when we model between-sample dependence, since it is not straightforward to introduce dependence in the discrete data among different units. To address these challenges, we propose a DAG model for mixed data with sample dependence and develop a de-correlation method that facilitates the structure learning of causal DAGs under these settings.

Although our method is broadly applicable across a range of data domains, the primary application in this work is the inference of gene regulatory networks (GRNs) from single-cell RNA sequencing (scRNA-seq) data. We model a GRN by a causal DAG, where a directed edge $V_i \rightarrow V_j$ implies that gene $i$ directly regulates the expression of gene $j$. Cells may originate from the same lineage or follow shared differentiation trajectories, leading to implicit association in gene expression patterns. Such dependency structures induce intrinsic correlations among cells, which may be further amplified by technical factors such as batch effects. As another example, intercellular signaling and spatial organization create structured dependence among cells \citep{almet2024inferring}, leading to correlated gene expression patterns. As a result, gene expression measurements in scRNA-seq data may violate the i.i.d. assumption underlying many standard causal discovery methods. Moreover, gene expression can be represented as a combination of continuous and discretized data, yielding a mixed data setting.

GRN inference methods have been developed specifically for single-cell applications. \citet{pratapa2020benchmarking} provide a comprehensive benchmark study of various GRN inference methods using synthetic GRNs and evaluates their performance across multiple criteria. Among the top performing methods, GENIE3 \citep{huynh2010inferring} performs GRN inference via a series of regression problems. Specifically, for each target gene, a tree-based ensemble model (e.g. random forest) is trained to predict its expression as a function of all other genes. Feature importance scores are aggregated across models to infer regulatory edges. Similarly, GRNBoost2 \citep{moerman2019grnboost2} builds upon GENIE3 by using gradient boosting machines to improve performance and scalability to large single-cell datasets. Both GENIE3 and GRNBoost2 rely on predictive models for the conditional distribution $[V_i|V_{-i}]$. As such, the resulting regulatory network captures statistical dependencies but lack a causal interpretation. Another top performing method, PIDC \citep{chan2017gene}, infers GRN structures via an information-theoretic method based on mutual information between gene pairs but only estimates an undirected network. While these methods have demonstrated empirical results in recovering plausible biological interactions in single-cell data, they all rely on the implicit assumption that individual cells are independent and identically distributed. As discussed above, cells may exhibit significant dependence with each other. These challenges motivate the development of our causal discovery method that explicitly accounts for sample dependence in a mix of multiple data types.

\subsection{Related work}

In this work, we focus on learning the underlying causal structure among a set of variables $V = (V_1, \ldots, V_p)$. We assume that data is generated from a DAG $\mathcal{G}$ whose structure encodes direct causal relationships. The set of parent nodes of variable $V_j$ is denoted by $PA_j \subseteq \{1,\ldots,p\} \setminus \{j\}$. Under standard i.i.d. assumptions, each sample $x_i = (x_{i1},\ldots,x_{ip})$ for $i = 1,\ldots,n$ is independently drawn from a joint distribution that factorizes according to the DAG,
\begin{align}
    f(V_1,\ldots,V_p) = \prod_{j = 1}^{p} f(V_j|V_{PA_j}).
\end{align}
Causal graph learning under i.i.d. assumptions has been well-explored in a variety of settings; see \citet{glymour2019review} and \citet{nogueira2022methods} for recent reviews.
In contrast, there are few methods that consider dependence among observational units $\{x_1,\ldots,x_n\}$. One related direction exists in the causal inference literature on \textit{partial interference}, where units may influence one another \citep{pmlr-v115-bhattacharya20a}. However, partial interference methods typically assume that the causal DAG over variables is known and focus on estimating unit-level causal effects.
Our goal is to recover the causal DAG $\mathcal{G}$ while allowing for dependence across observational units.

Let $X = (X_1,\ldots,X_p) \in \mathbb{R}^{n\times p}$ denote the observed data matrix, where each column $X_j = (x_{1j},\ldots,x_{nj})^\top \in \mathbb{R}^n$ is a vector of observations across all $n$ units. \citet{JMLR:v25:21-0846} proposed a linear Gaussian structural equation model on network data by introducing dependent exogenous variables $\varepsilon_j = (\varepsilon_{1j},\dots,\varepsilon_{nj})$:
\begin{align}
\begin{split}
    X_j = \sum_{k \in PA_j} \beta_{kj}X_k + \varepsilon_j, \quad \varepsilon_j \sim \mathcal{N}_n (0, \Sigma),
    \label{eq:model}
\end{split} 
\end{align}
where $\beta_{kj}$ is the direct causal effect of $X_k$ on $X_j$ along the directed edge $k\to j$ in the underlying DAG $\mathcal{G}$. The $n\times n$ covariance matrix $\Sigma=(\sigma_{ab})$ is positive definite and models the covariance between two units such that $\sigma_{ab}=\text{Cov}(\varepsilon_{aj},\varepsilon_{bj})$ for $j=1,\ldots,p$. Using the Cholesky factor $L$ of $\Theta = \Sigma^{-1}$ as a means to de-correlate the data matrix $X$ results in de-correlated data $\widetilde{X} = L^{\top}X$. Applying a standard causal discovery method on the de-correlated data $\widetilde{X}$ shows promising results for DAG learning. However, this approach is limited to continuous data and is not applicable to discrete variables. 
In addition, the support of $\Theta$ is restricted to a known network (graph) among the $n$ units, which is somewhat limited for real applications.

In our prior work \citep{pmlr-v258-chen25e}, we extend the framework of \citet{JMLR:v25:21-0846} by introducing a latent-variable formulation for binary data:
\begin{align}
    z_{ij} &= \sum_{k\in PA_j}\beta_{kj}x_{ik} + \varepsilon_{ij}=x_i \beta_j +\varepsilon_{ij}, \label{eq:cont} \\ 
    x_{ij} &= I(z_{ij} > 0), \label{eq:binary}
\end{align}
for $i \in [n] := \{1,\dots,n\}$ and $j \in [p] := \{1,\dots,p\}$, where $\beta_j=(\beta_{1j},\ldots,\beta_{pj})\in\mathbb{R}^p$ such that $\text{supp}(\beta_j)=PA_j$. To introduce between-unit dependence, the noise term $\varepsilon_j = (\varepsilon_{1j},\ldots,\varepsilon_{nj})$ is assumed to follow a multivariate normal distribution, $\varepsilon_j \sim \mathcal{N}_n(0,\Sigma)$, where $\Sigma$ captures the dependence among units. Under this latent-variable formulation, $x_{ij}$ corresponds to thresholding an unobserved latent variable $z_{ij}$. Since $x_{ij}$ is binary, one cannot simply transform them by $\wh{L}$, the Cholesky of an estimated $\wh{\Sigma}^{-1}$. Instead, the latent vectors $Z_j = (z_{1j},\ldots,z_{nj})^{\top}, j \in [p]$ are imputed by sampling from a multivariate truncated Gaussian distribution conditional on the observed binary outcomes. To achieve de-correlation, the imputed latent continuous variables $Z_j$ are de-correlated with $\wh{L}$. Then, any standard causal discovery method can be applied to the de-correlated $Z_j, j\in [p]$ to estimate a causal graph.

\subsection{Contributions of this work}

This paper makes two primary contributions. First, we apply the de-correlation framework for causal discovery to GRN learning from single-cell RNA-seq data with potential between-cell dependence. However, single-cell expression data may consist of both continuous gene expression and multi-level discrete states, which cannot be handled by existing de-correlation approaches \citep{JMLR:v25:21-0846,pmlr-v258-chen25e}. To address this, we generalize the latent-variable formulation in \eqref{eq:cont} and \eqref{eq:binary} and the associated de-correlation method to mix of general discrete and continuous variables. Second, we apply our proposed framework to an extensive study on inferring gene regulatory networks from scRNA-seq data in \citet{chu2016single}. We further develop a stability measure through a bootstrap method to quantify uncertainty in estimated gene regulatory interactions (edges in the learned graph). High-confidence edges indeed are supported by known biological interactions reported in the literature, demonstrating the effectiveness of our de-correlation approach for causal discovery in complex biological data.

The rest of the paper is structured as follows. Section~\ref{sec:model} introduces a new DAG model for dependent mixed data and presents an overview of de-correlation. In Section~\ref{pre_est_tau_sig}, we describe procedures for pre-estimating some model parameters, in particular, the covariance matrix among the units. Section~\ref{de-correlation_alg} develops the latent variable recovery and de-correlation method for mixed data. Section~\ref{simul_sec} presents simulation studies on both random and real DAGs. In Section~\ref{app_sec}, we apply our method to the single-cell data and evaluate our estimated gene regulatory relationships against findings reported in the biological literature. Section~\ref{sec:conc} concludes with a discussion of our main findings and potential directions for future research.

\section{Model and overall idea}\label{sec:model}

\subsection{DAG model for dependent mixed data}
\label{DAG_sec}

We generalize the SEM in Equations~\eqref{eq:cont} and \eqref{eq:binary} to mixed data, where each variable $X_j$ may be continuous, $x_{ij}\in\mathbb{R}$, or discrete with $C_j$ levels, $x_{ij}\in\{0,1,\ldots,C_j-1\}$. 
To construct a unified model for the two types of variables, a set of continuous latent variables $z_{ij}$, $i\in[n]$ and $j\in[p]$, is introduced to define the underlying SEMs and generate observed mixed data through transformations. 

Let $\mathcal{D} \subseteq [p]$ denote the index set of discrete variables and $\mathcal{C} = [p] \setminus \mathcal{D}$ the index set of continuous variables. For $j\in\mathcal{D}$, the discrete value of $x_{ij}$ is determined by $z_{ij}$ and a set of thresholds $T_j = \{\tau_{j,c}:\tau_{j,0} < \tau_{j,1} < \cdots < \tau_{j,C_j}\}$, where $\tau_{j,0} = -\infty$ and $\tau_{j,C_j} = \infty$. We define a quantization mapping of $z_{ij}$ given the thresholds $T_j$:
\begin{align}\label{eq:quant_mapping}
    Q(z_{ij}; T_j) = c 
    \quad \text{if} \quad 
    \tau_{j,c} < z_{ij} \le \tau_{j,c+1}, 
    \qquad c = 0,\ldots,C_j-1.
\end{align}
Let $\mathcal{G}$ be a DAG over $p$ nodes with parent sets $PA_j,j\in[p]$.
Our model for mixed data with between-unit dependency is defined by the following SEM associated with $\mathcal{G}$:
\begin{align} 
    z_{ij} &= \sum_{k\in PA_j}\beta_{kj}x_{ik} + \varepsilon_{ij}=x_i \beta_j +\varepsilon_{ij}, \quad \varepsilon_j \sim \mathcal{N}_n(0,\Sigma),\label{model:latent} \\  
    x_{ij} &= 
\begin{cases}
Q(z_{ij}; T_j), & \text{if } j\in \mathcal{D} \\
z_{ij}, & \text{if } j\in \mathcal{C}
\end{cases}, \label{eq:discrete}
\end{align}
for $i\in[n]$ and $j\in[p]$, where the coefficient vector $\beta_j=(\beta_{1j},\ldots,\beta_{pj})\in\mathbb{R}^p$ such that $\text{supp}(\beta_j)=PA_j$. Similarly to \citet{pmlr-v258-chen25e}, we introduce dependency across the rows of $X$ by assuming that the error vector $\varepsilon_j = (\varepsilon_{1j},\ldots,\varepsilon_{nj})$ follows a multivariate normal distribution in~\eqref{model:latent}. We fix $\text{diag}(\Sigma) = 1$ so that the parameters $\{\beta_j,T_j,\Sigma\}$ are identifiable. The error terms $\varepsilon_{ij}$ are regarded as exogenous variables, representing the source of randomness in the structural equation model. The dependence between exogenous noise $\varepsilon_{ij}$ across units induces dependence among the corresponding units $x_i$. This formulation allows unit-level dependence to be explicitly modeled through the covariance structure of the exogenous variables.

A possible alternative to the quantization mapping approach in Equations~\eqref{model:latent} and \eqref{eq:discrete} is to model each discrete variable $X_j$ as a multi-logit regression of its parents, as proposed in \citet{gu2019penalized}. However, multi-logit models introduce a large number of parameters: 
A $C$-categorical multi-logit model requires $C - 1$ sets of regression coefficients, one for each category other than the baseline, leading to a total of $(C - 1)p$ parameters for each discrete variable. Instead, our model has much fewer parameters for a discrete variable of a large number of levels. Moreover, the multi-logit formulation ignores any potential ordering among the discrete levels, treating them as categorical rather than ordinal. This limitation is particularly relevant in applications such as gene regulatory network inference, where discretized gene expression levels (e.g. low/medium/high) are ordinal by nature. Our model in Equations~\eqref{model:latent} and \eqref{eq:discrete} reflect this ordering, making it more suitable for discretization used in single-cell RNA-seq data analysis.

\subsection{Overall idea of de-correlation}\label{decor_subsec}

Under our model in \eqref{model:latent} and \eqref{eq:discrete}, both the continuous columns $X_{\mathcal{C}}$ and discrete columns $X_{\mathcal{D}}$ in the data $X$ are generated through the continuous latent data $Z =[Z_{\mathcal{C}}, Z_{\mathcal{D}}]= [X_{\mathcal{C}}, Z_{\mathcal{D}}]$, where $Z_j = \sum_{k \in PA_j} \beta_{kj}X_k + \varepsilon_j$ for $j\in[p]$. 
We can rewrite Equation~\eqref{model:latent} in terms of the latent variables: 
\begin{align}
    Z_j = \sum_{k \in PA_j} \beta_{kj}g(Z_k) + \varepsilon_j,
\qquad j \in [p], \label{eq:modelZ}
\end{align}
where $g(Z_k) = Z_k$ for continuous parents $k \in \mathcal{C}$ and $g(Z_k) = Q(Z_k; T_k)$ for discrete parents $k \in \mathcal{D}$. Equation~\eqref{eq:modelZ} defines an SEM over $Z = [Z_1,\ldots,Z_p]$ associated with the same DAG $\mathcal{G}$ over the observed data $(X)$. 
This observation motivates our proposal to learn the structure of $\mathcal{G}$ from the continuous latent data $Z$. 

However, there is between-sample dependency among $z_i$, the $i$th row of $Z$, $i\in[n]$, due to the dependent model for $\varepsilon$ in Equation~\eqref{model:latent}. De-correlation aims to remove the sample-level dependencies such that each unit contributes independent information, which is a key step in our procedure to accurately estimate causal graphs in the dependent data setting. 
Let $\Theta = \Sigma^{-1}$ be the precision matrix and $L$ be its Cholesky factor, i.e., a lower-triangular matrix such that
$\Theta = LL^{\top}$. Since
\begin{align*}
    \text{Cov}(L^{\top}\varepsilon_j) = L^{\top}\Sigma L = L^{\top}(LL^{\top})^{-1}L = I_n,
\end{align*}
the transformed error vector $\widetilde{\varepsilon}_j=L^{\top}\varepsilon_j$ has independent components $\widetilde{\varepsilon}_{ij},i\in[n]$. Let $e_i=(\varepsilon_{i1},\ldots,\varepsilon_{ip})$ be the vector of all error variables for the $i$th unit
and $\widetilde{e}_i=(\widetilde\varepsilon_{i1},\ldots,\widetilde\varepsilon_{ip})$.
The SEMs \eqref{model:latent} for $j=1,\ldots,p$ define a mapping $F:\mathbb{R}^p \mapsto \mathbb{R}^p$ such that $z_i=F(e_i)$.
Then, $\widetilde{z}_i=F(\widetilde{e}_i)$ becomes independent across the units $i\in[n]$, to which one can apply standard DAG learning
methods for causal discovery. This is the overall idea of de-correlation of the latent data $Z$. Two key steps are (1) to estimate the
covariance matrix~$\Sigma$; (2) to impute and de-correlate the error variables $\varepsilon_j$ and reconstruct the de-correlated latent data $\widetilde{Z}=[\widetilde{Z}_1,\ldots,\widetilde{Z}_p]$.

{Our method consists of two phases that we briefly summarize here with the detailed development provided in the next two sections.}
{Phase 1, which we call the pre-estimation phase, focuses on estimating the covariance matrix $\Sigma$ and the threshold values $T = (T_j)_{j\in \mathcal D}$. We maximize the likelihood of the discrete data to estimate $T$. Given $\widehat T$, we estimate the covariance matrix $\Sigma$ using a pairwise likelihood approach.}
{Phase 2 is the de-correlation phase. We develop an EM algorithm to iteratively reconstruct de-correlated latent data $\widetilde{Z}$ and update the SEM parameters $\beta_j,j\in[p]$ in Equation~\eqref{model:latent}, while fixing the estimates $\wh{\Sigma}$ and $\wh{T}$ obtained in Phase 1.
This procedure provides de-correlated, continuous data $\widetilde{Z}$ on which one may apply any standard DAG learning algorithm to recover the underlying causal graph. 

\section{Pre-estimation of $T$ and $\Sigma$}\label{pre_est_tau_sig}

In the pre-estimation phase, we obtain initial estimates of the threshold parameters $T = (T_j)_{j\in \mathcal D}$, regression coefficients $\beta=(\beta_{1},\ldots,\beta_p)$, and covariance matrix $\Sigma$. Our approach begins by estimating $(T,\beta)$ under the initialization of $\Sigma^{(0)} = I_n$, and then estimates $\Sigma$ conditional on the resulting $(\wh{T},\wh{\beta})$. 
Together, the two components yield estimates of $(\wh{T},\wh{\beta},\wh{\Sigma})$. During the de-correlation procedure in Section~\ref{de-correlation_alg}, $\wh{T}$ and $\wh{\Sigma}$ will be fixed while $\wh{\beta}$ will be used as the initial value for further update.

The pre-estimation phase requires an initial estimate of the parent set $PA_j$ for each node $j\in[p]$. We apply the mixed data version of the Max-Min Hill Climbing (MMHC) algorithm \citep{tsamardinos2006max} to learn a DAG structure from the observed data. The estimated parent sets, denoted by $\wh{PA}_j$ for $j\in[p]$, in the learned DAG are used as the initial estimates.

\subsection{Threshold estimation}\label{sec:threshold}

Let $\phi_n(\cdot;\mu,\Sigma)$ be the density of $\mathcal{N}(\mu,\Sigma)$.
Recall that $X\beta_j$ denotes the linear combination formed by the parent variables of node $j$ as $\text{supp}(\beta_j)=PA_j$. 
Our starting point is the likelihood of $X_j$, $j\in\mathcal D$ given $X\beta_j$ as specified by the SEMs in \eqref{model:latent} and \eqref{eq:discrete}:
\begin{align}\label{eq:likTSigma}
    P(X_{j}|X\beta_j; T_{j}, \Sigma) = \int_{D(X_j,T_j)} \phi_n(Z_j;X\beta_j,\Sigma)\, d\,Z_j,
\end{align}
where $D(X_j,T_j) = \prod_{i = 1}^n (\tau_{j,x_{ij}}, \tau_{j,x_{ij}+1}]$ is the Cartesian product of the truncation intervals implied by the observed levels $x_{ij}, i\in[n]$.

Joint estimation of $\beta$, $T$, and $\Sigma$ is extremely challenging due to the high dimensionality of $\Sigma$ and the complexity of the $n$-variate integral in \eqref{eq:likTSigma}. To address this in a practical way, we first estimate $\beta_j$ and $T_j$ while fixing $\Sigma=I_n$, which leads to two substantial simplifications: (1) the likelihood decomposes across samples, enabling efficient evaluation of the integral in Equation~\eqref{eq:likTSigma}, and (2) it decouples the estimation of $(T_j,\beta_j)$ for different discrete variables.
Specifically, we arrive at the following  log-likelihood for each $j\in\mathcal{D}$:
\begin{align}\label{eq:loglikelihood}
\ell(T_j,\beta_j \mid X)
= \sum_{i=1}^n \log\!\left[\Phi\!\left(\tau_{j,x_{ij}+1} - x_i \beta_j\right) - \Phi\!\left(\tau_{j,x_{ij}} - x_i \beta_j\right)\right],
\end{align}
where $\Phi(\cdot)$ is the cumulative distribution function of $\mathcal{N}(0,1)$.
We alternate between optimizing $T_j$ and $\beta_j$ until convergence as outlined in Algorithm~\ref{alg:taubetajoint}, which is a blockwise coordinate ascent algorithm to maximize the log-likelihood $\ell(T_j,\beta_j \mid X)$ in~\eqref{eq:loglikelihood}.
\begin{algorithm}[H]
\caption{Pre-estimation of $T_j$ and $\beta_j$ by coordinate ascent}
\label{alg:taubetajoint}
\begin{flushleft}
Input: $\wh{PA}_j$ and $T_j^{(0)}$

Iterate between the following steps for $t=1,2,\ldots$ until a stopping criterion is met:
\begin{enumerate}
    \item $\beta^{(t)}_j\gets \argmax_{\beta_j}\ell(T^{(t-1)}_j,\beta_j \mid X)$. 
    \item $T^{(t)}_j\gets \argmax_{T_j}\ell(T_j,\beta^{(t)}_j \mid X)$. 
\end{enumerate}
\end{flushleft}
\end{algorithm}
\noindent
The initial thresholds, $T_j^{(0)} = (\tau_{j,1}^{(0)},\ldots,\tau_{j,C_j-1}^{(0)})$, are chosen by the empirical distribution of the observed levels. Specifically, we compute the frequency $\widehat{p}_{j,c}$ of each level in the data $X_j$ and set,
\begin{align*}
    \tau_{j,c}^{(0)} = \Phi^{-1}\left( \sum_{k = 1}^{c} \widehat{p}_{j,k} \right),
    \quad c = 1,\ldots, C_j - 1,
\end{align*}
where $\Phi^{-1}(\cdot)$ is the quantile function of $\mathcal{N}(0,1)$.

Since each discrete observation $x_{ij}$, $j\in\mathcal{D}$ is obtained by thresholding a latent Gaussian variable $z_{ij}$, the likelihood of the observed data involves integrals over multivariate truncated Gaussian distributions. In particular, for each variable $j$, the latent vector $Z_j = (z_{1j},\ldots,z_{nj})^{\top}$ follows a multivariate Gaussian distribution with constraints on $Z_j$ imposed by the observed discrete outcomes. As a result, the observed data likelihood does not admit a closed-form maximizer with respect to $\beta_j$ for fixed thresholds $T_j$. To address this, we develop an EM algorithm~\citep{dempster1977maximum} to maximize the observed data likelihood $\ell(T^{(t-1)}_j,\beta_j \mid X)$ over $\beta_j$ in Step~1. In the E-step, we compute the conditional expectation of each $z_{ij}$ given the observed level $x_{ij}$, which is the mean of a truncated normal distribution. In the M-step, we perform a linear regression of the expectation of $Z_j$ on the parent variables $X_{\widehat{PA}_j}$ to update $\beta_j$. We perform a small fixed number of iterations, 
which guarantees monotone increase of the observed data likelihood in \eqref{eq:loglikelihood} at each step while substantially reducing computational cost.

In Step 2, we estimate the thresholds vector $T_j = (\tau_{j,1},\ldots,\tau_{j,C_j-1})$ by maximizing the log-likelihood in \eqref{eq:loglikelihood} subject to $\tau_{j,1}<\dots<\tau_{j,C_j-1}$. To enforce this constraint, we re-parameterize the thresholds in terms of their successive differences: 
\begin{align*}
    \delta_{j,1} = \tau_{j,1}, \quad \delta_{j,c} = \tau_{j,c} -\tau_{j,c-1}, \quad c = 2,\ldots,C_j-1,
\end{align*}
where $\delta_{j,c} >0$ for $c \geq 2$. The resulting optimization problem for $\delta_j$ is solved numerically through the L-BFGS-B algorithm, resulting in $\wh{T}_j$.
Empirical convergence of Algorithm~\ref{alg:taubetajoint} appears to be very fast, as shown in the Supplementary Material.

\subsection{Covariance estimation}
\label{cov_est_sec}

Fixing $\wh{T}$ and $\wh{\beta}$, we estimate the covariance matrix $\Sigma$ to capture the dependence among units. Recall that $\text{diag}(\Sigma) = 1$. Due to the high-dimension of $\Sigma$, we develop a pairwise approach to estimate off-diagonal correlation $\rho_{ab}$ for each pair of units $a$ and $b$, by integrating both discrete and continuous variables.

Without loss of generality, let us consider the estimation of $\rho_{12}$, the correlation between the first two units, $x_1$ and $x_2$. As $\text{diag}(\Sigma) = 1$, the two exogenous error variables, $(\varepsilon_{1j}, \varepsilon_{2j})$, follow a bivariate Gaussian distribution,
\begin{align}
\begin{pmatrix}\varepsilon_{1j}\\ \label{eq:biv_dist_eq}
\varepsilon_{2j}
\end{pmatrix} &\sim  \mathcal{N}
\begin{bmatrix}
\begin{pmatrix}
0\\
0\\
\end{pmatrix}\!\!,&
\begin{pmatrix}
1 & \rho_{12}\\
\rho_{12} & 1\\
\end{pmatrix}
\end{bmatrix} \hspace{0.5em} \text{for all } j \in [p].
\end{align}
There are two types of unit pairs among $\{(x_{1j},x_{2j}), j\in[p]\}$, pairs of discrete values and pairs of continuous values. Based on Equation~\eqref{eq:discrete}, a pair of discrete values implies the following constraints on $(\varepsilon_{1j},\varepsilon_{2j})$:
\begin{align*}
x_{1j} &= c_1 \quad \Longleftrightarrow \quad \tau_{j,c_1} < \varepsilon_{1j} + x_1 \beta_j \leq \tau_{j,c_1+1},  \\ \
x_{2j} &= c_2 \quad \Longleftrightarrow \quad \tau_{j,c_2} < \varepsilon_{2j} + x_2 \beta_j \leq \tau_{j,c_2+1}.
\end{align*}
Given regression coefficient $\beta_j$, we use 
Equation~\eqref{model:latent} to find the probability mass function of $(x_{1j}, x_{2j})$ through the CDF of a bivariate Gaussian:
\begin{align}\label{eq:probabilitydiscrete}
    P(x_{1j}, x_{2j}|D_j^{1,2}, \rho_{12}) = \iint\limits_{D_j^{1,2}} \phi \left(u_1,u_2|\rho_{12} \right)du_1 du_2,
\end{align}
where $\phi(\cdot,\cdot|\rho_{12})$ is the density function of the bivariate Gaussian distribution in \eqref{eq:biv_dist_eq} and $D_j^{1,2} \subset \mathbb{R}^2$ is the domain of the integral. If $(x_{1j}, x_{2j}) = (c_1,c_2)$, then the domain $D_j^{1,2} = (\tau_{j,c_1} - x_{1}\beta_j, \tau_{j,c_1+1} - x_1\beta_j] \times (\tau_{j,c_2} - x_2\beta_j, \tau_{j,c_2+1} - x_2\beta_j]$.
For a continuous variable pair, their joint density is
\begin{align}\label{eq:probabilitycontinuous}
    p(x_{1j}, x_{2j} \mid \rho_{12}) &= \phi(x_{1j} - x_1\beta_j,\; x_{2j} - x_2\beta_j \mid \rho_{12}).
\end{align}

To estimate $\rho_{12}$, we construct a pairwise likelihood that incorporates both discrete and continuous pairs across all $p$ variables:
\begin{align}\label{alg:likelihood_cov_mat}
L(\rho_{12} \mid x_1, x_2)
&= \prod_{j\in \mathcal{D}} P(x_{1j}, x_{2j} \mid D_j^{1,2}, \rho_{12})
   \times \prod_{j\in \mathcal{C}} p(x_{1j}, x_{2j} \mid \rho_{12}).
\end{align}
For each pair of units $(a,b)\in [n]\times [n]$, we estimate the correlation $\rho_{ab}$ by maximizing the likelihood function~\eqref{alg:likelihood_cov_mat}. This uni-variate optimization can be efficiently solved using standard numerical methods.

Following the setting of \citet{JMLR:v25:21-0846}, we assume a block-diagonal structure for $\Sigma$, where each block captures dependence within a group of units and units from two different blocks are independent. This type of structure naturally arises in settings such as social networks, cell populations sharing a common lineage, or spatially localized units. Sparsity may exist within individual blocks, but we do not impose this assumption in our simulations or empirical analysis. 

To estimate the block-diagonal covariance matrix among the units, we find the pairwise maximum likelihood estimate of each correlation $\rho_{ab}$ between two units in the same block. After estimating all within-block correlations for each block, we check whether the resulting covariance matrix $\wh{\Sigma}$ is positive definite to ensure $\wh{\Theta} = \wh{\Sigma}^{-1}$ is well defined. If not, negative eigenvalues of $\wh{\Sigma}$ are truncated to zero and off-diagonal elements are rescaled by a factor $<1$ (e.g. 0.9) to improve numerical stability.

\section{De-correlation for structure learning}\label{de-correlation_alg}

\subsection{Latent data recovery and de-correlation}

In the de-correlation phase, we fix $\Sigma$ and $T$ to their pre-estimated values, $\wh{\Sigma}$ and $\wh{T}$. Let $\wh{L}$ be the Cholesky factor of $\wh{\Theta} = \wh{\Sigma}^{-1}$. We separate the de-correlation procedure between continuous variables and discrete variables. Continuous variables are de-correlated with $\wh{L}$ by $\widetilde{X}_j = \wh{L}^{\top}X_j$, $j \in \mathcal{C}$. For discrete variables, we estimate the latent data $Z_{\mathcal{D}}$ and coefficients $\beta_j$, $j\in\mathcal{D}$ via an EM algorithm~\citep{dempster1977maximum}. Given $Z_{\mathcal{D}}$, we apply the Cholesky factor $\wh{L}$ to generate de-correlated data $\widetilde{Z}_j = \wh{L}^{\top}Z_j$, $j \in \mathcal{D}$. The full de-correlated dataset is $\widetilde{Z} = [\widetilde{X}_{\mathcal{C}}, \widetilde{Z}_{\mathcal{D}}]$, in which all variables are continuous.

In the E-step, we impute the latent data by computing its expectation from a truncated Gaussian distribution given the observed discrete outcomes and the current estimated coefficients $\wh{\beta}$. 
Recall that the distribution of the error vector is $\varepsilon_j \sim \mathcal{N}_n(0,\Sigma)$ from \eqref{model:latent}. Conditional on the observed data $X$, however, $\varepsilon_j = (\varepsilon_{ij}, i\in[n])$ must fall within the intervals implied by $T_j$: 
\begin{align}\label{eq:condeps}
    \tau_{j,x_{ij}} - x_i\beta_j < \varepsilon_{ij} \leq \tau_{j,x_{ij}+1} - x_i\beta_j, \quad i\in[n].
\end{align}
This corresponds to sampling $\varepsilon_j$ from a $n$-dimensional truncated Gaussian distribution, $\mathcal{N}_{T}(0,\Sigma; T_j)$. 

When $n$ is large, directly sampling from an $n$-dimensional truncated Gaussian distribution is computationally challenging. To address this, we exploit the block-diagonal structure of $\Sigma$, which limits dependence to within-block units. We develop a Gibbs sampler to generate samples within each block. Averaging over $N$ draws yields an approximate expectation $\mathbb{E}(\varepsilon_j\mid X,\beta_j)$ to reconstruct the latent continuous data $\wh{Z}_{\mathcal{D}}$ via \eqref{model:latent}.
 
The recovered latent data, $\wh{Z}_{\mathcal{D}}$, can then be de-correlated to generate $\widetilde{Z}_j = \wh{L}^{\top}\wh{Z}_j$, similar to the continuous variables. Then, the M-step updates the regression parameters $\beta_j$ by regressing $\widetilde{Z}_j = \wh{L}^{\top}Z_j$ on $\wh{L}^{\top}X$ for $j\in\mathcal{D}$, which now satisfies, approximately, the assumption of independent errors. Ridge regularization is applied to avoid overfitting.

Algorithm~\ref{alg:discdag} details the iterative process of recovering the latent data $Z_j$ for discrete variables, de-correlating $Z$, and estimating $\beta$. Our de-correlation approach may be understood under a unified perspective for continuous and discrete variables,
\begin{align}\label{eq:decorr}
    \widetilde{Z}_j = \wh{L}^{\top}\wh{Z}_j &= \wh{L}^{\top}X\wh{\beta}_j + \wh{L}^{\top}\wh{\varepsilon}_j, \quad j\in[p],
\end{align}
where $\wh{Z}_j = X_j$ and $\widetilde{Z}_j = \widetilde{X}_j$ for $j \in \mathcal{C}$. When $\wh{\Sigma}$ is well estimated, $\wh{L}^{\top}\wh{\varepsilon}_j$ effectively removes unit-level dependence, yielding approximately independent units. This de-correlation transformation is the key step in improving causal structure learning on dependent  data.

\begin{algorithm}[ht]
\caption{EM-based de-correlation for mixed data}
\label{alg:discdag}
\begin{flushleft}
Input: Mixed data $X = [X_{\mathcal{C}}, X_{\mathcal{D}}]$, $\wh \Sigma$, $\wh{T}$, $\wh{\beta}^{(0)}$, and $\wh{PA}_j, j\in[p]$.

Let $\widetilde{X}_j = \wh{L}^{\top}X_j$, $j\in\mathcal{C}$.

Iterate the following steps for $t = 1,2,\ldots$ until a stopping criterion is met:
\begin{enumerate}
    \item For each discrete variable $j \in \mathcal{D}$:
    \begin{enumerate}
        \item Draw samples of $\varepsilon_j$ from $\mathcal{N}_T(0,\widehat{\Sigma}; \wh{T}_j)$ by a Gibbs sampler.
        \item Approximate the expectation $\wh{\varepsilon}^{(t)}_j = \mathbb{E}[\varepsilon_j \mid X, \widehat{\beta}^{(t-1)}_j]$ via Monte Carlo average of $\varepsilon_j$.
        \item Reconstruct latent variables $\widehat{Z}^{(t)}_j = X \widehat{\beta}^{(t-1)}_j + \widehat{\varepsilon}^{(t)}_j$.
    \end{enumerate}
    
    Form the de-correlated latent data $\widetilde{Z}^{(t)}_{\mathcal{D}}=\wh{L}^{\top} \wh{Z}^{(t)}_{\mathcal{D}}$ and let $\widetilde{Z}^{(t)}=[\widetilde{X}_{\mathcal{C}},\widetilde{Z}^{(t)}_{\mathcal{D}}]$.
    
    \item For $j \in \mathcal{D}$, regress $\widetilde{Z}^{(t)}_j$ on $\widehat{L}^{\top} X_{\widehat{PA}_j}$ to update $\widehat{\beta}^{(t)}_j$.
\end{enumerate}
\end{flushleft}
\end{algorithm}

\subsection{Structure learning}\label{sec:structure_learn}

Standard structure learning methods for i.i.d. continuous data may now be applied on the de-correlated data, $\widetilde{Z}^{(t)}$ recovered in the final iteration of Algorithm~\ref{alg:discdag}, to infer the underlying DAG. However, the conditional expectation of the latent variables $Z_{\mathcal{D}}$, is obtained via Monte Carlo sampling in the E-step. To reduce variability, we aggregate information across multiple iterations of $\widetilde{Z}^{(t)}$ by pooling structure learning results via a consensus vote.

In general, a causal DAG is not identifiable from observational data alone, since multiple DAGs may encode the same set conditional independence relations. These DAGs are called Markov equivalent. A Markov equivalence class can be uniquely represented as a completed partially directed acyclic graph (CPDAG). Thus, our goal is to estimate a CPDAG from observational data. We consider two strategies for obtaining the final CPDAG estimate. In the \textit{Consensus} approach, we run a standard structure learning algorithm (such as MMHC) on $M$ de-correlated datasets $\{\widetilde{Z}^{(t)}: t\in S$, $|S| = M\}$, where $S$ is a subset of iterations from Algorithm~\ref{alg:discdag}. An edge is accepted into the final causal graph estimate if it appears in at least half of the estimated CPDAGs. The second strategy is called \textit{Average}, where we first compute an element-wise average of the $M$ de-correlated datasets and then perform the same standard causal discovery method on the averaged data.

From a computational perspective, the average approach requires a single graph estimation whereas the consensus approach requires $M$ estimates of a causal graph. The consensus approach can yield more stable edge recovery but for large variable sizes, the average approach may be more applicable. We benchmark both methods against a \textit{Baseline} approach, which applies the mixed data version of the same  structure learning method directly to the original data. We first show the improvement of using our method on simulated data, and then move to the motivating application of inferring gene regulatory networks from single-cell RNA-seq data.

\section{Simulation results}\label{simul_sec}
For simulated data, we use synthetic random DAGs and real DAGs from the \textit{bnlearn} repository \citep{scutari2017bayesian} that are widely used as benchmark networks for evaluating structure learning algorithms. We compare the true CPDAG to the estimated CPDAG learned from our approach. To compare methods, we compute the number of true positives (TP), false positives (FP), and false negatives (FN). According to the interpretation of undirected edges in a CPDAG, we regard an undirected edge $i - j$ as two directed edges $i \rightarrow j$ and $j \rightarrow i$. A directed edge counts as a true positive if it appears in both the learned and true CPDAGs, false positives correspond to directed edges present only in the learned CPDAG, and false negatives correspond to directed edges present only in the true CPDAG. Under this metric, incorrect edge orientations are penalized and undirected edges are accounted for in both directions. We summarize the performance of CPDAG recovery using the F1-score,
\begin{equation}
    \text{F1-score} = \frac{2 \times \text{Precision} \times \text{Recall}}{\text{Precision} + \text{Recall}},
\end{equation} 
where precision is defined as $\text{TP}/(\text{TP}+\text{FP})$ and recall as $\text{TP}/(\text{TP}+\text{FN})$. Note the F1-score ranges between $0$ and $1$, and a value of $1$ means perfect recovery.

\subsection{Setup}
For each DAG structure, edge weights were sampled independently and uniformly from the interval $[-0.9,-0.6] \cup [0.6,0.9]$. Exogenous noise terms were then generated according to Equation~\eqref{model:latent} with a specified covariance structure $\Sigma$ (detailed below). Given sampled exogenous noise terms and edge weights, latent continuous variable $Z_j$ was generated through \eqref{model:latent}. With probability $0.5$, $Z_j$ was discretized by the quantization mapping in \eqref{eq:discrete} with thresholds $\tau_{j,1} = -1$ and $\tau_{j,2} = 1$. We repeated this process for $j = 1,\ldots,p$ according to a topological ordering of the DAG, which produced a mixed data matrix $X \in \mathbb{R}^{n \times p}$. 

The covariance structure $\Sigma$ followed a block-diagonal structure, where each block corresponded to a cluster of correlated units of size 10 to 15. For each block, we randomly chose between \textit{Equal} and \textit{Toeplitz} covariance patterns. 
\begin{itemize}
    \item \textit{Equal} structure emulates fully-connected units, where $\Sigma_{ij} = \theta$ if $i \neq j$ with $\theta \sim \mathcal{U}(0.4, 0.7)$. This occurs when all units in the block are equally dependent with one another.
    \item \textit{Toeplitz} structure emulates units connected in a Markov chain, where $\Sigma_{ij} = \theta^{|i - j|/5}$ and $\theta \sim \mathcal{U}(0.1,0.25)$.
\end{itemize}
These covariance designs allow us to emulate varying levels and types of inter-unit dependence across clusters.

\subsection{Evaluation of covariance estimation}\label{sec:cov_pre_est}

We evaluated the accuracy of covariance matrix estimation as it plays an essential role for effective de-correlation in Algorithm~\ref{alg:discdag}. We simulated mixed data according to ten random DAGs, with $n \in\{100, 500\}$ units and $p \in \{100, 500\}$ variables for each DAG. Using the threshold estimates by Algorithm~\ref{alg:taubetajoint}, we applied the pairwise MLE method from Section~\ref{cov_est_sec} to estimate each correlation $\wh{\rho}_{ij}$ within the known block structure and computed the element-wise RMSE between the estimated covariance matrix $\wh{\Sigma} = (\wh{\rho}_{ij})_{n\times n}$ and the true covariance matrix $\Sigma = (\rho_{ij})_{n\times n}$:
\begin{align*}
    \text{RMSE} (\wh \Sigma , \Sigma) = \left\{ \frac{1}{|H|}\sum_{(i,j)\in H} (\wh \rho_{ij} - \rho_{ij})^2\right\}^{1/2},
\end{align*}
where $H$ is the set of non-zero, off-diagonal elements in $\Sigma$. 

We conducted two simulation studies to evaluate $\text{RMSE} (\wh \Sigma , \Sigma)$. In the first simulation, we assumed knowledge of the true parent sets. 
In the second simulation, we repeated the same experiment but used parent sets estimated by MMHC on the original mixed data. To quantify the accuracy of parent recovery, we calculated the F1-score of the estimated parent sets. For $n = 500$ and $p = 500$, the average F1-score was $0.75$. For smaller sample sizes ($n = 100$, $p = 500$), parent recovery was substantially less accurate with the average F1-score of $0.41$.
Figure~\ref{fig:cov_est} shows that the results from using the estimated parents and the true parents were similar for all settings. This indicates that the proposed covariance estimation procedure remains robust when the estimated graph is inaccurate. Additionally, covariance estimation improves as $p$ increases, since the likelihood-based estimation in \eqref{alg:likelihood_cov_mat} effectively treats the number of variables as the sample size.

\begin{figure}
    \centering
    \includegraphics[width=0.5\linewidth]{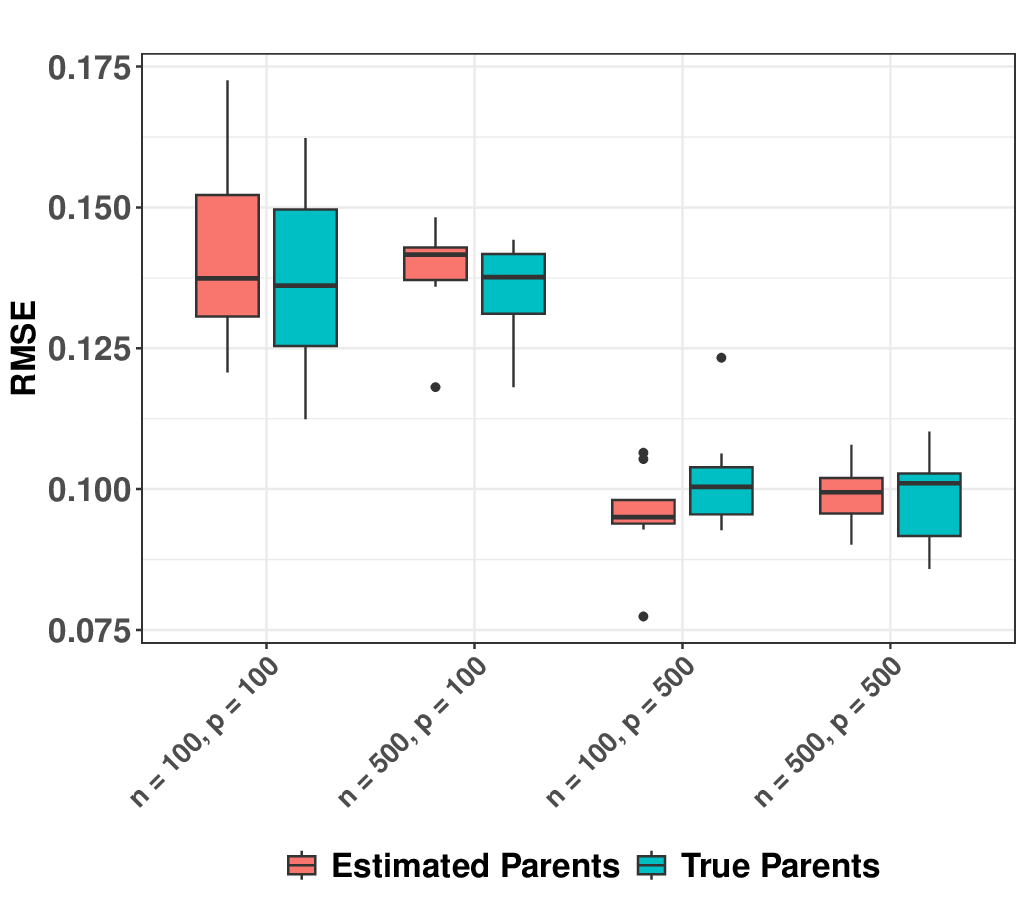}
    \caption{RMSE of covariance estimate across data size settings of $n \in \{100, 500\}$ and $p \in \{100, 500\}$ using either estimated parent sets from MMHC on the mixed data or true parent sets.}
    \label{fig:cov_est}
\end{figure}

To further assess the accuracy of the covariance estimates and the effectiveness of the de-correlation procedure, we examined pairwise correlations across observational units. We only use the subset of continuous variables, as correlation measures for discrete data are less straightforward to define and interpret. Using simulated datasets of size $(n,p) = (500,500)$, we compared the correlations within the blocks of the original data matrix $X_j,j\in \mathcal{C}$ with those of the de-correlated variables $\widetilde{X}_j,j\in \mathcal{C}$.
As shown in Figure~\ref{fig:cov_before_after}, the original data exhibit strong correlations across units, with an average of 0.62, indicating that the data obviously violates the independence assumption commonly required for causal discovery algorithms. After applying our de-correlation method, the distribution of correlations is concentrated around zero with an average of 0.09. This demonstrates that the transformation indeed substantially removed cross-unit dependence and yielded approximately uncorrelated units.

\begin{figure}
    \centering
    \includegraphics[width=0.5\linewidth]{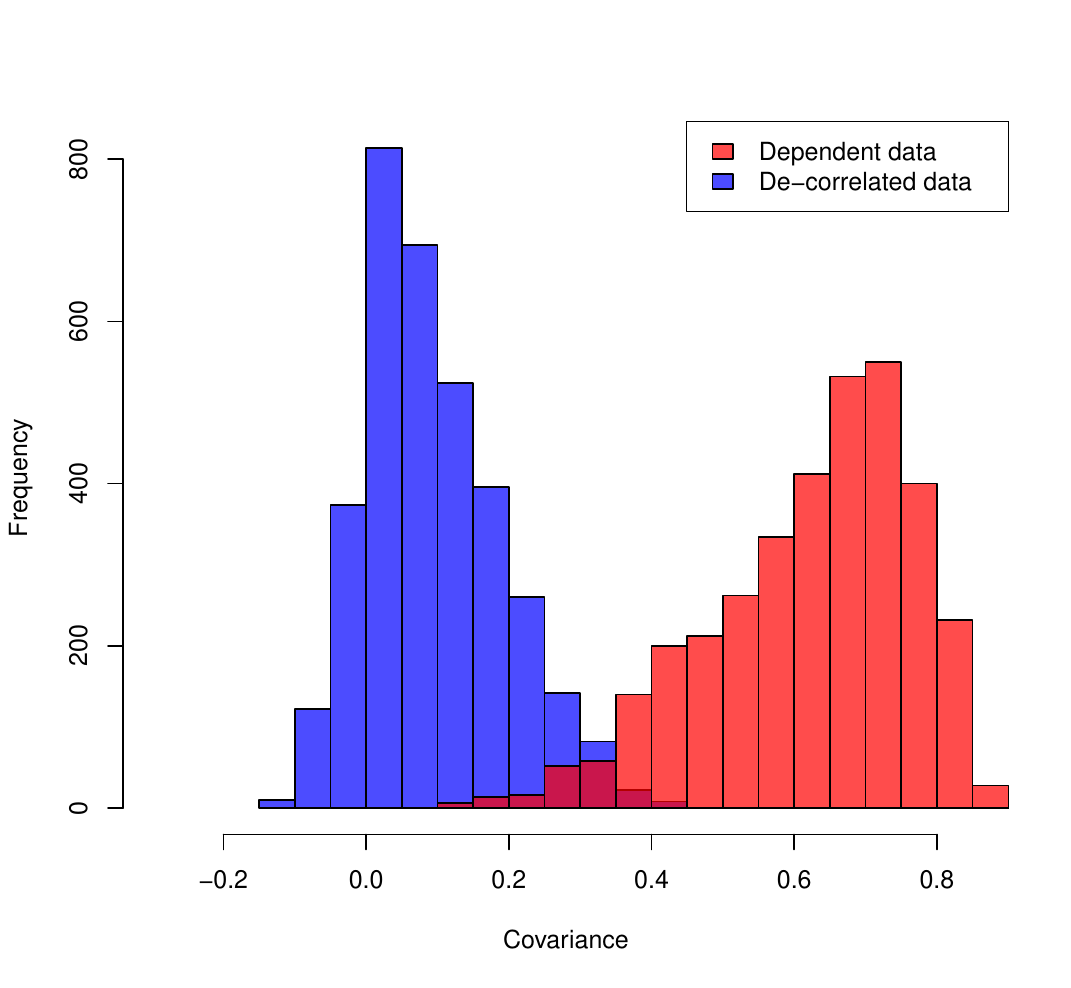}
    \caption{Histograms of the within block correlations among the continuous variables of the original data $X_j,j\in\mathcal{C}$ and the corresponding de-correlated $\widetilde{X}_j,j\in\mathcal{C}$.}
    \label{fig:cov_before_after}
\end{figure}

\subsection{DAG learning accuracy}

To demonstrate consistency of our de-correlation algorithm across causal discovery methods, we used three different DAG learning methods: MMHC~\citep{tsamardinos2006max}, PC, and Copula-PC~\citep{cui2016copula}. MMHC is a hybrid causal discovery algorithm that first identifies potential parent sets using conditional independence tests and then orients edges using a hill climbing algorithm to optimize a scoring function. PC is a constraint-based method that estimates the skeleton of a DAG using conditional independence tests and then orients edges by a set of orientation rules known as the Meek's rules \citep{10.5555/2074158.2074204}. Copula-PC is an extension from the PC algorithm that assumes the data follows a Gaussian copula model.  
It uses Gibbs sampling to draw covariance matrices from a posterior distribution. The samples are used to compute an average correlation matrix and an effective sample size, which are input to the standard PC algorithm for causal discovery. For each DAG learning method, we use the mixed data version on the dependent mixed data (baseline) and the continuous version on the de-correlated data from our algorithm (average and consensus approaches). We then compare the results to quantify the performance of de-correlation. We simulated 10 DAGs in each parameter setting in our experiments below. 

We first evaluated the performance of our method on mixed data simulated from random DAGs with $n \in \{100, 500\}$ units, $p \in \{100,1000\}$ variables and $2p$ edges. As reported in Figure~\ref{fig:sim_data_f1}, the consensus and average approaches show significant improvement over the baseline approach in all experiments. The magnitude of improvement varies between settings with $p = 100$ and $p = 1000$. As discussed in Section~\ref{sec:cov_pre_est}, covariance estimation improves as the number of variables increases. Consequently, the increased performance gains observed when $p = 1000$ are attributable to more accurate estimation of $\Sigma$, which leads to more effective de-correlation. Compared to MMHC and PC, Copula-PC had the worst performance in the baseline approach, but the magnitude of improvement from the baseline to consensus approach is the largest. Because Copula PC relies on sampling correlation matrices, the large variable size makes accurate sampling difficult. The consensus approach explicitly addresses sample dependence and discretization to improve causal graph estimation with Copula PC. Regardless of the DAG learning method used, we observe improved causal structure recovery via our de-correlation algorithm.
\begin{figure}
    \centering
    \includegraphics[width=0.8\linewidth]{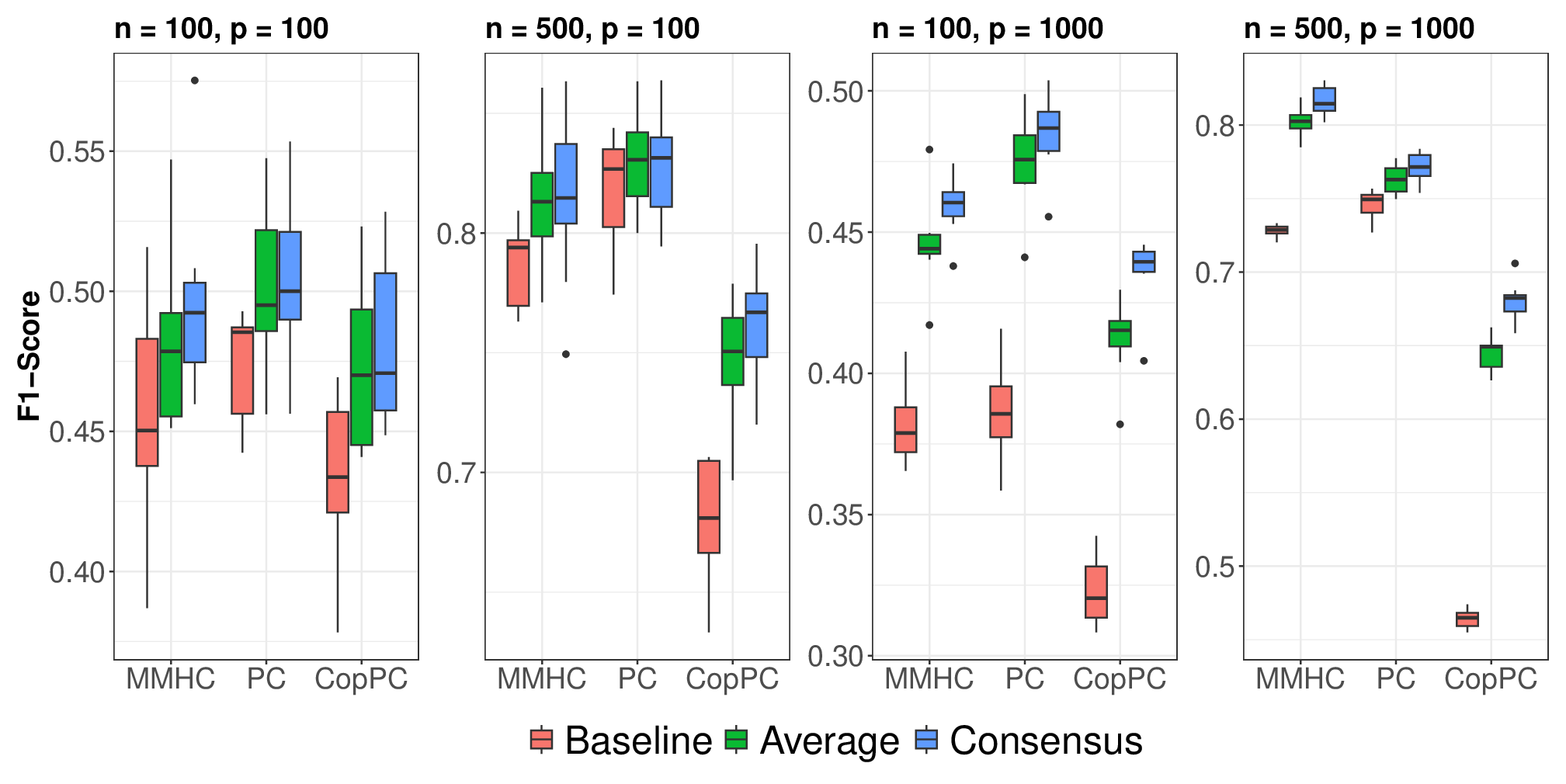}
    \caption{F1-scores of MMHC, PC, and Copula-PC algorithms on the original mixed data (baseline) and de-correlated continuous data (average and consensus) for $n = 100, 500$ and $p = 100, 1000$.}
    \label{fig:sim_data_f1}
\end{figure}

Next, we evaluated our method using eight benchmark DAGs ($p$ variables and $s$ edges) available in the \textit{bnlearn} repository~\citep{scutari2017bayesian}: Hailfinder ($p = 56, s = 66$), Hepar2 ($p = 70, s = 123$), Win95pts ($p = 76, s = 112$), Munin1 ($p = 186, s = 273$), Andes ($p = 223, s = 338$), Pigs ($p = 441, s = 592$), Diabetes ($p = 413, s = 602$), Link ($p = 724, s = 1125$). We simulated mixed data from each DAG and repeated the process 10 times. To reflect varying data dimensionalities, smaller networks, Hailfinder, Hepar2, Win95pts, and Munin1, were simulated with sample sizes $n \in \{100,200\}$. For larger and more complex networks, Andes, Pigs, Diabetes, and Link, we used sample sizes $n \in \{100,500\}$. This design provided a range of experimental conditions: both low-dimensional ($n > p$) and high-dimensional ($p > n$) settings, allowing us to assess the performance under different sample-to-variable ratios. Figure~\ref{fig:bn_results} summarizes the F1-scores for DAG recovery across the eight networks. For nearly all networks and sample sizes, the de-correlation-based methods, the average and consensus approaches, consistently outperformed the baseline approach. 
\begin{figure}
    \centering
    \includegraphics[width=.9\linewidth]{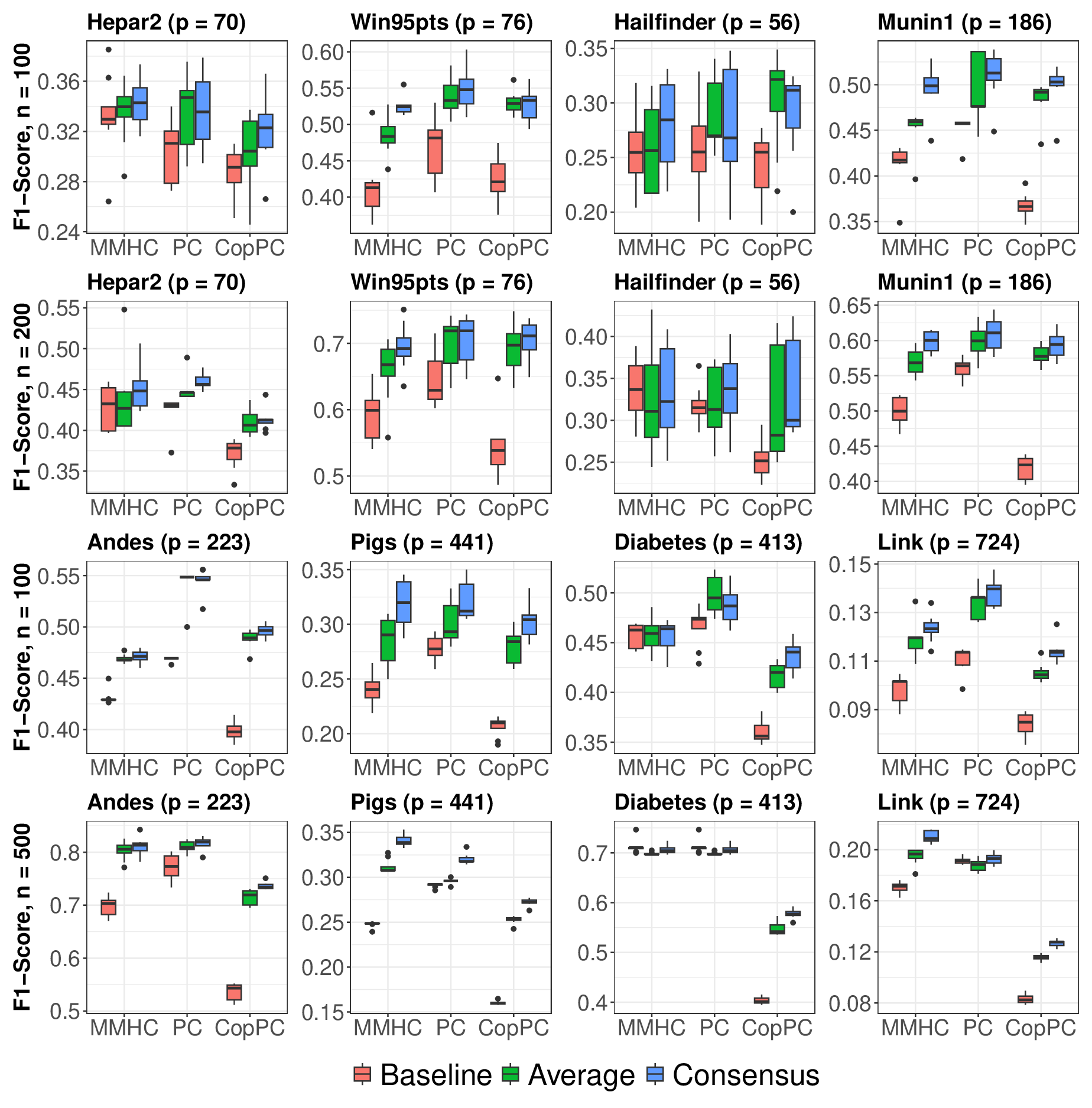}
\caption{F1-scores of MMHC, PC, and Copula-PC algorithms on original mixed data (baseline) and de-correlated continuous data (average and consensus) for four smaller networks (Hepar2, Win95pts, Hailfinder, Munin1) with $n = 100, 200$ and four larger networks (Andes, Pigs, Diabetes, Link) with $n = 100, 500$.}
\label{fig:bn_results}
\end{figure}

In experiments with smaller networks such as Hepar2 and Hailfinder, we observe modest but consistent improvements over the baseline approach with $n = 100$ across MMHC, PC, and Copula PC. With $n = 200$, the F1-scores for these networks using MMHC are comparable across the baseline, average, and consensus approaches. However, using PC and Copula PC yields higher F1-scores for the consensus and average approaches than the baseline approach. For Win95pts and Munin1, the average and consensus approaches show greater improvements over the baseline for both $n = 100$ and $n = 200$ across all causal discovery methods. For example, with $n = 100$, the Munin1 network using the MMHC method sees a $20\%$ increase in F1-score by the consensus approach over the baseline.  

For larger graphs with limited sample size ($n = 100$), including Andes, Pigs, and Link, the consensus and average approaches achieve substantially higher F1-scores than the baseline across causal discovery methods. In the Diabetes network with $n = 100$, MMHC yields similar performance across all approaches, while PC and Copula PC show improved F1-scores for the consensus and average approaches relative to the baseline. This pattern largely persists at $n = 500$. For Andes and Pigs with $n = 500$, the consensus and average approaches consistently outperform the baseline across all methods. We observe a $38\%$ and $16\%$ increase in F1-score comparing the baseline and consensus approaches using the MMHC method for Pigs and Andes, respectively. For Diabetes, MMHC and PC show comparable performance across approaches, whereas Copula PC, again, sees the most improvement through the consensus and average approaches. In the Link network with $n = 500$, PC yields similar F1-scores across all approaches, while MMHC and Copula PC show improved performance using the consensus and average approaches compared to the baseline. We observe a $24\%$ increase in F1-score of the consensus approach over the baseline approach for the MMHC method on the Link network with $n = 500$.

Overall, the consensus approach consistently achieves the best or comparable DAG recovery in almost all cases, while the average approach provides a computationally efficient alternative with comparable DAG recovery.

\section{Application to GRN inference}
\label{app_sec}

Gene regulatory networks describe how genes regulate each other through activation or repression, providing an understanding of how cells function and differentiate. Uncovering the causal relationships between genes is critical for understanding complex biological processes and how their dysregulation leads to diseases. In therapeutic settings, identifying which genes to up-regulate or inhibit can reveal potential targets for improving treatment efficacy. Because it is expensive and difficult to directly measure gene regulation, inferring GRNs from single-cell expression data is an active and important research area.

\subsection{Background and the data}

\citet{chu2016single} generated single-cell RNA-seq data to investigate how human embryonic stem cells (ESCs) transition from a \textit{pluripotent} state to \textit{lineage-specific progenitor} states. Pluripotent stem cells are uncommitted cells capable of differentiating into all three germ layers, ectoderm, mesoderm, and endoderm, while lineage-specific progenitors represent cells that have begun to develop towards a particular state. The key motivation of their study was to identify key transcriptional regulators that drive these transitions and to characterize gene regulation governing differentiation.

This dataset is publicly available at the Gene Expression Omnibus under accession number GSE75748. The dataset consists of 1,018 single cells spanning multiple cell types derived from ESCs and controls. These included 6 different cell types (H1s, $n = 212$; H9s, $n = 162$; DEC, $n = 138$; EC, $n = 105$; NPC, $n = 173$; TB, $n = 69$) along with human foreskins fibroblasts (HFFs; $n = 159$) that served as controls. There were 20,000 genes in the dataset but we focused on learning a GRN among $p = 51$ genes compiled by~\citet{chu2016single} that are known to be important lineage-informative marker genes. The genes we do not consider for GRN estimation are referred to as background genes.


\subsection{Pre-processing}\label{sec:pre-process}

Each expression profile was pre-processed following the normalization and quality control pipeline described in~\citet{li2018modeling}. In our analysis, we excluded HFFs (Human Foreskin Fibroblasts; controls) to focus on the developmental trajectory from pluripotent embryonic stem cells to lineage-specific progenitors. Excluding HFF cells, the data consists of $n = 859$ cells (units) and $p = 51$ genes (variables).

The raw measurements of single-cell RNA-seq data are counts of sequencing reads corresponding to transcripts for each gene in a cell. Due to technical noise, dropouts, and data sparsity, reads can be unreliable and highly variable. Additionally, many genes in scRNA-seq data exhibit discrete expression dynamics: Signaling genes may switch ``on'' only when a pathway is activated, while other genes may display multi-modal expression patterns. As a result, discretization of such genes can lead to more robust estimation of a causal GRN. To determine whether each gene should be modeled as continuous or discretized, we performed an additional test after the quality control pipeline. Specifically, for each gene $j$ with expression data $X_j = (x_{ij})_{i = 1}^n$, we fit both a single Gaussian model, 
\begin{align*}
    x_{ij} \sim \mathcal{N}(\mu_j,\sigma^2_j), \quad i = 1,\ldots,n,
\end{align*}
and Gaussian mixture models (GMMs) with $K = 2$ and $K = 3$ components,
\begin{align*}
    x_{ij} \sim \sum_{k = 1}^K \pi_{jk}\mathcal{N}(\mu_{jk}, \sigma^2_{jk}), \quad \text{where} \quad \sum_{k = 1}^K \pi_{jk} = 1, \quad \pi_{jk} > 0.
\end{align*}
We calculate the Bayesian Information Criterion (BIC) for each candidate model, and select the model with the smallest BIC value. If a Gaussian mixture model provides a better fit than the single Gaussian model, we further test the normality of each identified component using the Shapiro-Wilk test ($\alpha = 0.05$). If any mixture component deviates significantly from normality, we discretize the gene based on the GMM cluster assignments across $x_{ij},i\in[n]$. Conversely, if the single Gaussian best fits the data or if all mixture components pass the normality test, we keep the gene as a continuous variable. After the test for discretization, our dataset $X = (x_{ij})_{859 \times 51}$ has $|\mathcal{D}| = 20$ discrete  and $|\mathcal{C}| = 31$ continuous variables.

The distinct cell types in the scRNA-seq data suggest a block structure among cells since we expect cells of the same type to be more strongly correlated. To infer this structure from the RNA-seq data, we applied hierarchical clustering to the 859 cells and 2,000 randomly selected background genes as the feature vector. Figure~\ref{fig:cluster_dendro} shows that the dendrogram approximately partitions cells of the same type in the same clusters. Note that H1 and H9 cells, ESCs of the male and the female, cluster closely together due to their similar gene expression profiles \citep{sperger2003gene}. This verifies that it is reasonable to use the background genes to estimate the block structure among the cells.
\begin{figure}
    \includegraphics[width=0.65\linewidth, trim = 0 0.5in 0 0.7in, clip]{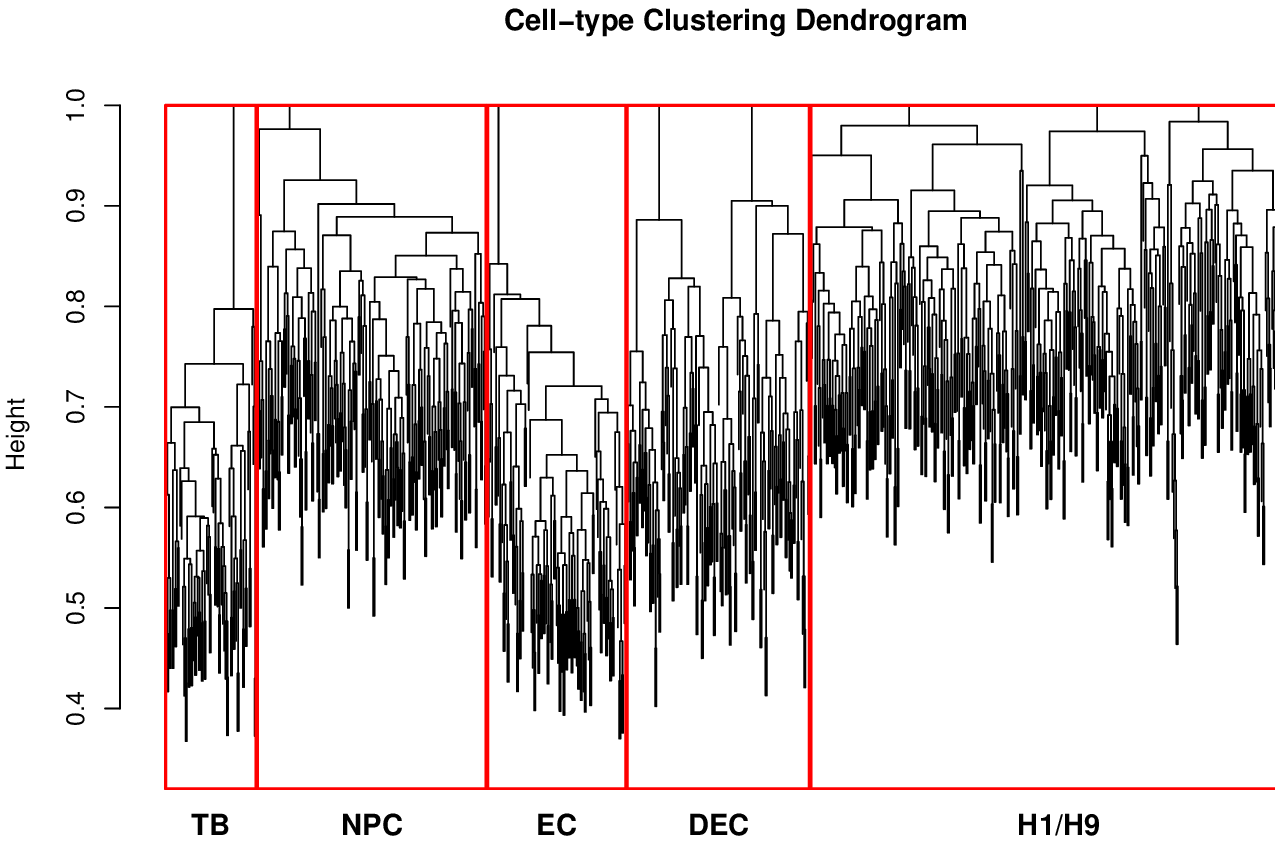}
    \caption{Hierarchical clustering dendrogram of cells with the approximate cell types corresponding to the clusters.}
    \label{fig:cluster_dendro}
\end{figure}

Considering the computation time and robustness of the covariance estimation method, we wanted to identify relatively small blocks of cells exhibiting high correlations. This consideration motivated our choice of using 100 clusters from the dendrogram. As shown in Figure~S4 in the Supplementary Material, $90\%$ of the clusters had size $\leq 30$.}

\subsection{Model Evaluation}

Without a true underlying GRN, we evaluated different GRN learning methods using cross-validation. Because likelihood evaluation requires independence across folds, we aggregated independent blocks of cells into a total of 10 folds. Each fold consisted of approximately $85$ cells. Let $\{X_{\text{train}}^{(k)}, X_{\text{test}}^{(k)}\}^{10}_{k=1}$ denote the 10-fold cross-validation splits of the scRNA-seq dataset.

We compared three different approaches to estimating the causal graph on each fold $k$: \textit{Baseline}, \textit{Consensus-Ident}, and \textit{Consensus}. 
\begin{itemize}
    \item Baseline: We apply a mixed data version of MMHC directly on $X_{\text{train}}^{(k)}$, the mixed-type gene expression data.
    \item Consensus: We apply Algorithm~\ref{alg:discdag} to $X_{\text{train}}^{(k)}$ and apply a continuous data version of MMHC on the resulting de-correlated continuous data $\widetilde{Z}_{\text{train}}^{(k)}$.
    \item Consensus-Ident: Identical to the consensus approach but we fix $\wh{\Sigma} = I_n$ in Algorithm~\ref{alg:discdag}, effectively removing the de-correlation step. We apply a continuous data version of MMHC on the dependent continuous data ${Z}_{\text{train}}^{(k)}$.
\end{itemize}
For each $k$, we applied each of the above methods on $X^{(k)}_{\text{train}}$ to obtain an estimated graph $\wh{\mathcal{G}}^{(k)}$ and the associated parameters $\wh\Psi^{(k)} := (\wh\beta^{(k)},\wh T^{(k)})$ under our model.

Test-data likelihood evaluation requires an estimate of the covariance matrix $\Sigma$. However, directly estimating $\Sigma$ from the held-out test data $X_{\text{test}}^{(k)}$ would lead to bias and inflate the resulting likelihood. On the other hand, we cannot use the training data to estimate the covariance for test data because the dependence among units varies across folds. To address these issues, we estimated the covariance structure using a randomly sampled set of 100 background genes. For each fold $k$, let $X_{\text{bg}}^{(k)} \in \mathbb{R}^{n_k \times 100}$ denote the expression matrix of 100 randomly sampled background genes measured on the $n_k$ cells of fold $k$. We applied the covariance estimation method as in Section~\ref{cov_est_sec} on $X_{\text{bg}}^{(k)}$ to estimate $\wh{\Sigma}^{(k)}_{\text{test}}$. Note that the test data $X^{(k)}_{\text{test}}$ was not used in the estimate of $\wh{\Sigma}^{(k)}_{\text{test}}$.

\begin{figure}
    \centering
    \includegraphics[width=0.5\linewidth]{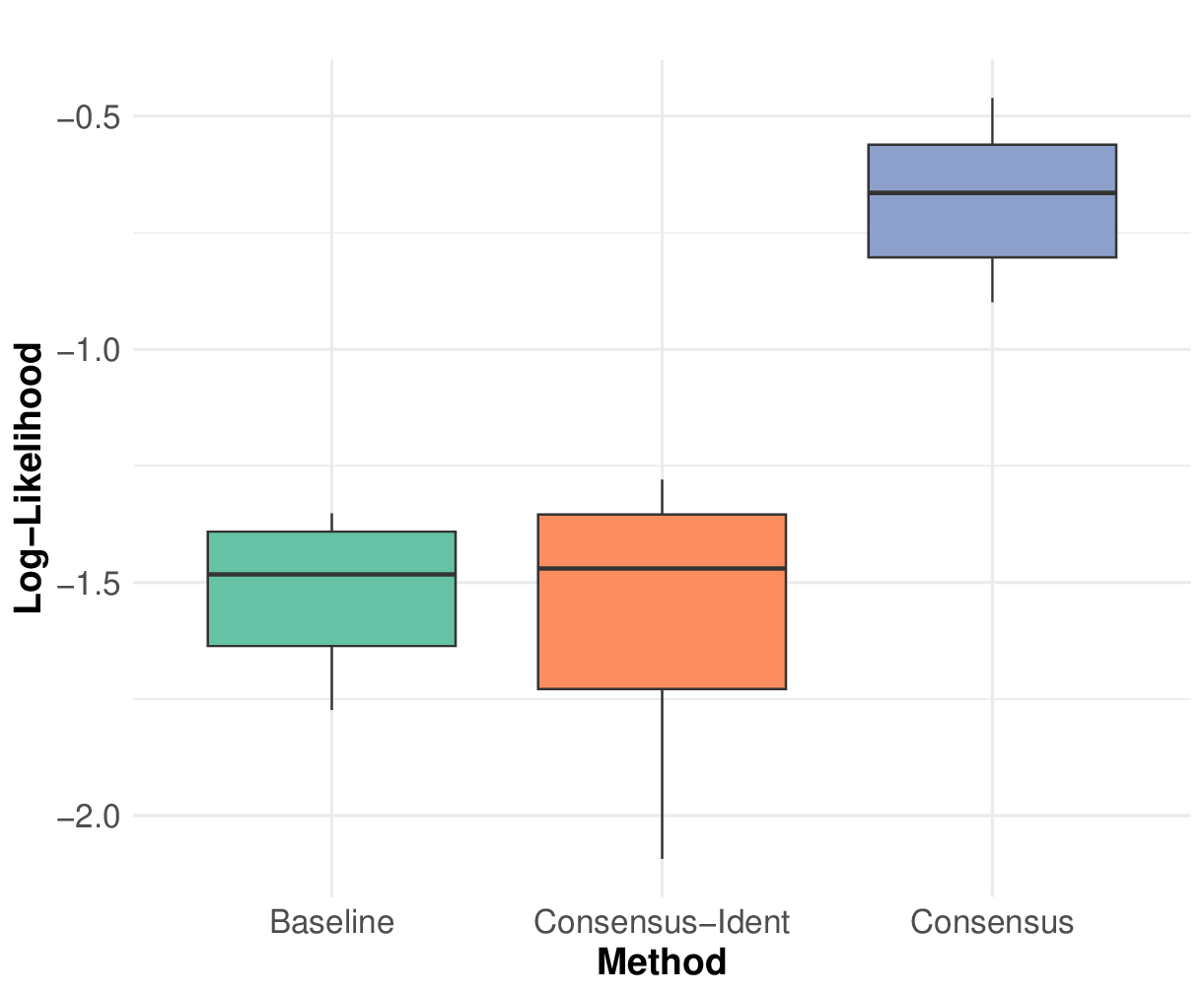}
    \caption{Boxplot comparison of test data log-likelihood distributions in cross validation across the baseline, consensus-ident, and consensus methods on single-cell data.}
    \label{fig:log_like_per_datapoint_mixed}
\end{figure}

Given $\wh{\mathcal{G}}^{(k)}$, $\wh{\Psi}^{(k)}$, and $\wh{\Sigma}^{(k)}_{\text{test}}$, we evaluated the log-likelihood of the test data $X_{\text{test}}^{(k)}$ under our mixed data model. For each fold, we report the normalized log-likelihood per data point, 
\[
\ell^{(k)} = \frac{1}{|X_{\text{test}}^{(k)}|} \log p(X_{\text{test}}^{(k)}|\wh{\mathcal{G}}^{(k)},\wh{\Psi}^{(k)},\wh\Sigma^{(k)}_{\text{test}}),
\]
where $|X_{\text{test}}^{(k)}|$ is the number of cells in the corresponding test dataset of fold $k$. This metric is a normalized log-likelihood to assess the model fit of each method accounting for fold size ensuring comparability across folds.
Figure~\ref{fig:log_like_per_datapoint_mixed} compares the normalized test log-likelihood among the three methods. The consensus method had a substantially higher median normalized log-likelihood ($-0.55$) than the baseline and consensus-ident approaches, which had a median of $-1.5$. Additionally, the distribution of test log-likelihoods for the consensus method does not overlap with those of the other two methods, indicating consistent and significant improvement in model fit.

The gap between the consensus and consensus-ident methods isolates the effect of de-correlation: Both methods use the same consensus approach and graph aggregation procedure, but only the consensus method explicitly removes sample dependence. The improved likelihood under the consensus method highlights the importance of de-correlating dependent units prior to causal discovery.

Overall, these results demonstrate that our proposed model provides a better fit to this scRNA-seq dataset. More broadly, it confirms that explicitly modeling dependence among cells can lead to significant improvement in model fitting.

\subsection{Analysis of Results}

Using the full dataset of $n = 859$ cells, we applied our consensus approach to obtain a final estimated CPDAG $\wh{\mathcal{G}}_{\text{full}}$. Recall that a CPDAG represents a Markov equivalence class and contains both directed ($\rightarrow$) and undirected edges ($-$). 

To quantify the robustness of each inferred regulatory interaction, we performed a resampling-based bootstrap analysis. In each bootstrap iteration $t = 1,\ldots,B$, we sampled 50\% of the cells ($429$ cells) with replacement and re-estimated the causal graph using the consensus method given in Algorithm~\ref{alg:discdag}. Because the block structure may vary across subsets of the data, we re-estimated the block structure for each bootstrap sample $t$. This procedure yielded a collection of estimated CPDAGs $\{\wh{\mathcal{G}}^{(t)}\}^{B}_{t = 1}$.

For each ordered pair $(a,b)$, let
\begin{align*}
    I^{(t)}_{a \to b}
    =
    \begin{cases}
    1, & \text{if } a \to b \in \wh{\mathcal{G}}^{(t)}, \\
    0.5, & \text{if } a - b \in \wh{\mathcal{G}}^{(t)}, \\
    0, & \text{otherwise}.
    \end{cases}
\end{align*}
Note that undirected edges contribute equally to both possible orientations. 

\begin{figure}
    \includegraphics[width=0.75\linewidth, trim = 0 0.6in 0 0.52in, clip]{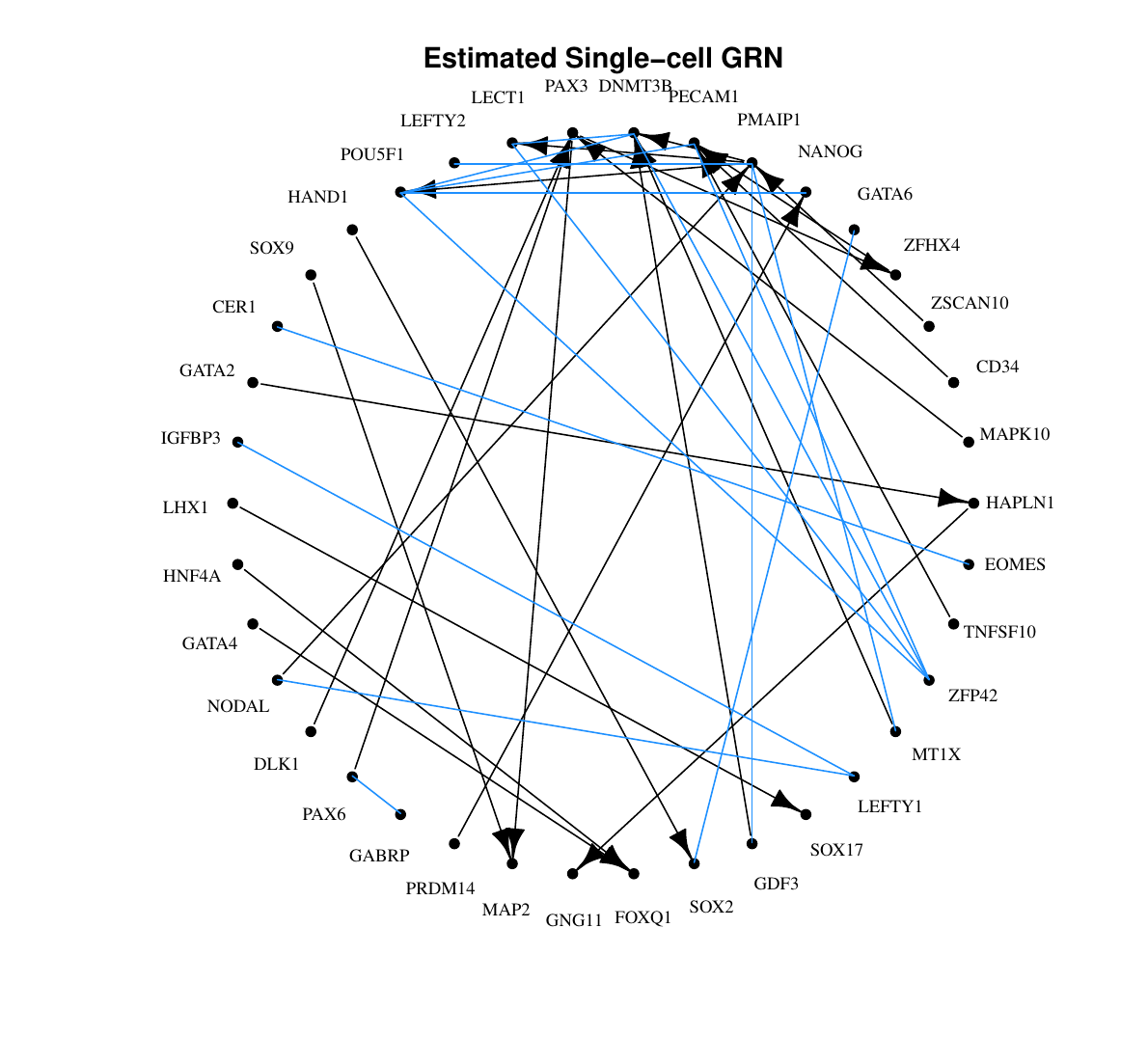}
    \caption{Gene regulatory network showing the predicted edges found using the bootstrap resampling analysis. Blue edges correspond to undirected edges and black edges correspond to directed edges.}
    \label{fig:grn_network}
\end{figure}

We aggregated $I^{(t)}_{a \to b}$ across the bootstrap graphs and accumulated counts for edges that appear in the full-data estimate $\wh{\mathcal{G}}_{\text{full}}$. For a directed edge $a \rightarrow b \in \wh{\mathcal{G}}_{\text{full}}$, we define its confidence score as
\begin{align*}
    \text{Conf}(a \rightarrow b) = \frac{1}{B}\sum_{t = 1}^{B} I^{(t)}_{a\rightarrow b}.
\end{align*}
Since an undirected edge $a-b$ in a CPDAG means that both orientations are possible, we define its confidence score as
\begin{align*}
    \text{Conf}(a-b) = \text{Conf}(a\rightarrow b) +  \text{Conf}(b \rightarrow a).
\end{align*}
\noindent
We retained all directed edges with $\text{Conf}(a \rightarrow b) \geq 0.5$ and undirected edges with $\text{Conf}(a - b) \geq 0.5$ in the final graph. For an undirected edge $a - b \in \wh{\mathcal{G}}_{\text{full}}$, we further compare the confidence for the two orientations. Specifically, if
\begin{align}
    \frac{\text{Conf}(a\rightarrow b)}{\text{Conf}(b\rightarrow a)} \geq 3 \quad \text{ or } \quad \frac{\text{Conf}(b\rightarrow a)}{\text{Conf}(a\rightarrow b)} \geq 3,
    \label{directional_support}
\end{align}
we orient the edge in the dominant direction. Otherwise, the edge remains undirected in the final graph. The resulting graph, $\wh{\mathcal{G}}_{\text{cf}}$, contains 39 edges, among which 23 are directed and 16 are undirected. The resulting network is shown in Figure~\ref{fig:grn_network}, where genes (nodes) without any edge connections are removed.

Table~\ref{table:select_lit_genes} presents the subset of predicted edges with either $\text{Conf}(a \rightarrow b) \geq 60\%$ or $\text{Conf}(a - b) \geq 80\%$ in the final graph and the literature support for some of the predicted edges. Top-ranked undirected edges capture tightly coupled regulators within the pluripotency network. For example, interactions among POU5F1, NANOG, and ZFP42 \citep{wang2006protein, mitsui2003homeoprotein}, as well as the feedback loop relationship between NODAL and LEFTY1 \citep{juan2001roles}, are known regulatory relationships. Several directed edges also align with known regulatory mechanisms, such as PRDM14 stabilizing NANOG-mediated pluripotency \citep{ma2011sequence}. DNMT3B and POU5F1 also have an experimentally established regulatory link by which DNMT3B suppresses POU5F1 transcription \citep{liu2025dna}, although our method estimated this as an undirected connection. Edges without direct literature support may present unknown causal interactions or arise from indirect regulatory pathways rather than direct effects. The full list of edges in our estimated GRN is provided the Supplementary Material.

\begin{table}
\centering
\caption{List of predicted edges with high-confidence in the inferred gene regulatory network}
\label{parset}
\begin{tabular}{lll}
\hline
$X_i - X_j$ & Confidence \% & Reference \\
\hline
POU5F1 $-$ ZFP42  & 100.0 & \citet{wang2006protein}\\
NANOG $-$ POU5F1 & 99.0  & \citet{mitsui2003homeoprotein}\\
LEFTY1 $-$ NODAL  & 98.5 & \citet{juan2001roles}\\
PECAM1 $-$ POU5F1 & 98.5  \\
PECAM1 $-$ ZFP42  & 97.5 \\
DNMT3B $-$ ZFP42  & 96.5 \\
DNMT3B $-$ LECT1  & 91.0  \\
CER1   $-$ EOMES  & 84.5 \\
DNMT3B $-$ POU5F1 & 84.5 & \citet{liu2025dna}\\
NODAL  $\rightarrow$ PMAIP1 & 73.0 \\
PAX3   $\rightarrow$ MAP2   & 72.5 \\
PAX6   $\rightarrow$ PAX3   & 71.3 \\
HAPLN1 $\rightarrow$ GNG11  & 66.5 \\
HNF4A  $\rightarrow$ FOXQ1  & 66.9 \\
GDF3   $\rightarrow$ DNMT3B & 66.3 \\
DLK1   $\rightarrow$ PAX3   & 64.5  \\
TNFSF10 $\rightarrow$ PECAM1 & 62.0  \\
PRDM14 $\rightarrow$ NANOG  & 61.2 & \citet{ma2011sequence}\\
MAPK10 $\rightarrow$ PAX3   & 61.0 \\
CD34   $\rightarrow$ PECAM1 & 61.0  \\
HAND1  $\rightarrow$ SOX2   & 60.0  \\
\hline
\end{tabular}
\label{table:select_lit_genes}
\end{table}

\section{Discussion and conclusion}
\label{sec:conc}
\subsection{Summary}
In this work, we introduced a principled framework for causal discovery on dependent mixed-type observational data. To model the data, we formulated a latent-variable SEM in which every observed variable is generated from an underlying continuous latent process. Discrete variables arise through a thresholding mechanism applied to latent Gaussian variables, while continuous variables are observed directly. A central feature of the proposed model is the cross-unit dependence via correlated exogenous noise terms. This induces dependence among samples that is orthogonal to causal relationships encoded by a DAG, which operates across variables. The two sources of dependence, one across units and one across variables, complicate traditional causal discovery methods that assume independence across units.

This motivates our de-correlation approach that leverages an estimated covariance matrix to transform the latent continuous data. After de-correlation, our samples are approximately independent while preserving the underlying causal structure among variables. Given the de-correlated data, we may use any standard causal discovery method to estimate the underlying DAG. A key insight is that causal discovery algorithms do not need to be redesigned for dependent data, but rather the data dependence can be addressed upstream through an appropriate transformation.

Across simulated and real networks, our method consistently resulted in better estimated causal graphs compared to  graphs estimated from the original data. Importantly, we showed that the improvement from our proposed framework is primarily due to the de-correlation of mixed data and not limited to a specific class of causal discovery algorithms. The method is particularly effective in high-dimensional settings with $p > n$ and strong dependence among units.

In the application to GRN learning from scRNA-seq data, we designed a bootstrap resampling method to quantify the stability of the inferred causal relationships between genes and predicted high-confidence interactions that aligned with  biological literature. We also showed through cross-validation that our method substantially outperformed direct application of causal discovery methods to single-cell data without de-correlation, supporting the notion of between-cell dependence in scRNA-seq data.

\subsection{Future work}
We outline directions for future work. First, our method is designed for observational data and does not incorporate experimental interventions or known causal constraints. In many real-world applications, partial interventional data or prior knowledge is available and can substantially improve causal identifiability and estimation accuracy. Extending our framework to integrate interventional data or incorporate prior causal knowledge would enhance its applicability in practical settings. Second, our data generating model and estimation procedure assume a linear structural equation model on the latent continuous data. While linear SEMs are widely used, they may be insufficient for capturing complex, potentially nonlinear interactions that arise in real-world systems.

Generalizing our method to nonlinear SEMs is feasible but introduces several challenges. Suppose we replace Equation~\eqref{model:latent} with 
\begin{align*}
    x_{ij} = f_{j}(x_{i,PA_j}, \varepsilon_{ij}),
\end{align*}
where $f_{j}(\cdot)$ is an unknown nonlinear function. Without loss of generality, assuming the variables of $X$ are sorted according to a topological order of the underlying DAG, $X_j$ can be written as a function of the noise variables up to index $j$,
\begin{align*}
    X_j = F_j(\varepsilon_1,\ldots,\varepsilon_j).
\end{align*}
If we can apply a Cholesky de-correlation to the noise terms, then the transformed variables can be expressed as
\begin{align*}
    \widetilde{X}_j = F_j(\widetilde{\varepsilon}_1,\ldots,\widetilde{\varepsilon}_j),
\end{align*}
where $\widetilde{\varepsilon}_j = L^{\top}\varepsilon_j$ are independent. In principle, any nonlinear causal discovery method designed for independent data could be applied to $\{\widetilde{X}_j,j\in[p]\}$. Adapting our current framework to the nonlinear setting is a promising important direction for future work.

\begin{funding}
This work was supported in part by NIH grant R01GM163245 and NSF grant DMS-2305631 (to QZ).
\end{funding}

\begin{supplement}
\stitle{Web appendices} \sdescription{Appendices containing (a) convergence diagnostics for Algorithm~\ref{alg:taubetajoint}, (b) additional simulations on estimation accuracy of thresholds, (c) additional results and experiments for scRNA-seq application.}
\end{supplement}



\bibliographystyle{imsart-nameyear} 
\bibliography{ref}       


\end{document}


\begin{frontmatter}
\title{Supplement to ``Causal Discovery on Dependent Mixed Data with Applications to Gene Regulatory Network Inference''}
\runtitle{Causal Discovery on Dependent Mixed Data}

\begin{aug}
\author[A]{\fnms{Alex}~\snm{Chen}}
\and
\author[A]{\fnms{Qing}~\snm{Zhou}}
\address[A]{Department of Statistics and Data Science,
University of California, Los Angeles}
\end{aug}

\end{frontmatter}

\section{Convergence diagnostics for Algorithm 1}
\begin{figure}[htbp]
    \centering
    \includegraphics[width=0.6\linewidth]{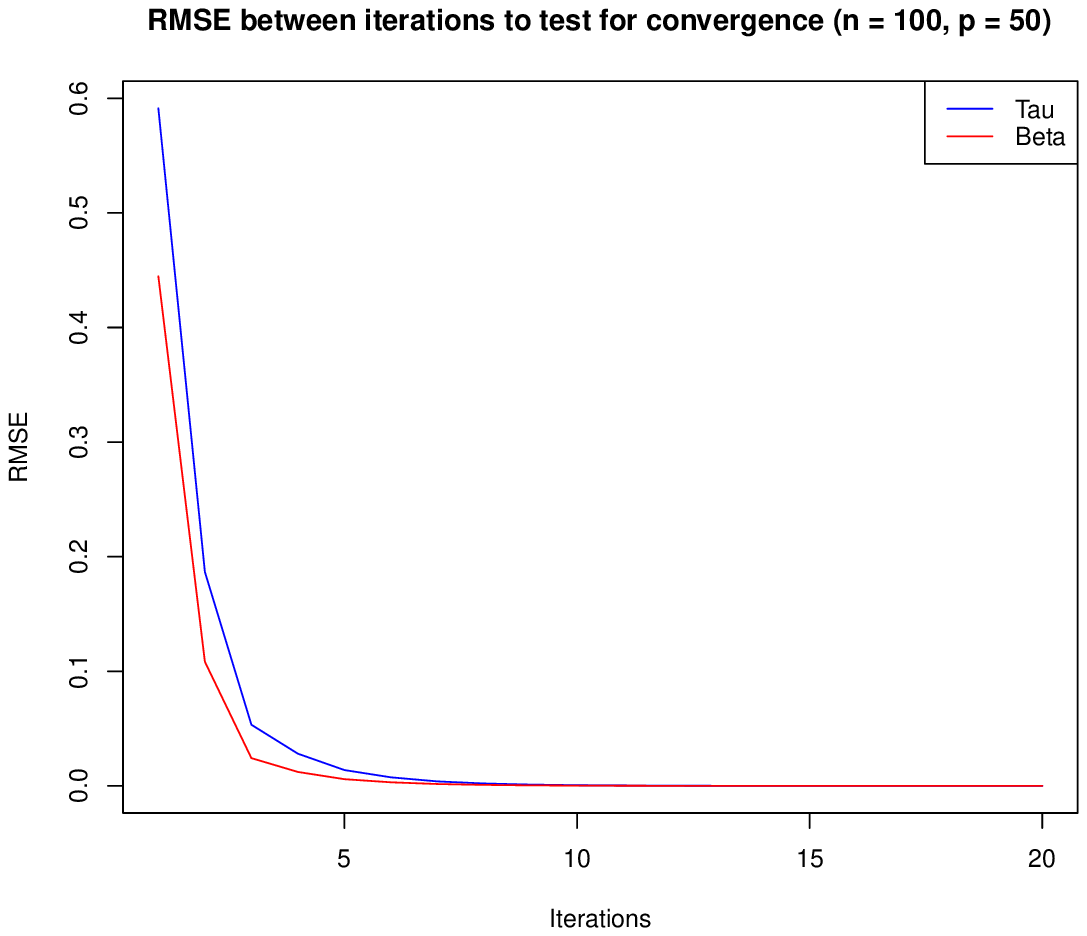}
    \caption{Difference between consecutive iterations of Algorithm 1 under the settings of $n = 100$ and $p = 50$ using a random DAG.}
    \label{fig:taubetaconverge}
\end{figure}

We assessed a normalized distance between iterations of thresholds $T$ and regression coefficients $\beta$. We ran a simulation under a random DAG with $n = 100$ units and $p = 50$ variables. For each iteration $t$, we computed
\begin{align*}
    d_{\beta}^{(t)} = \|\beta^{(t)} - \beta^{(t-1)}\|_2, \quad d_{T}^{(t)} = \|T^{(t)} - T^{(t-1)}\|_2.
\end{align*}
\noindent
The convergence behavior is illustrated in Figure~\ref{fig:taubetaconverge}, which shows that  $d_{\beta}^{(t)}$ and $d_{\tau}^{(t)}$ converged to $\approx 0$ after approximately 5 iterations.

\section{Threshold Estimation}
\begin{figure}[htbp]
    \centering
    \includegraphics[width=0.6\linewidth]{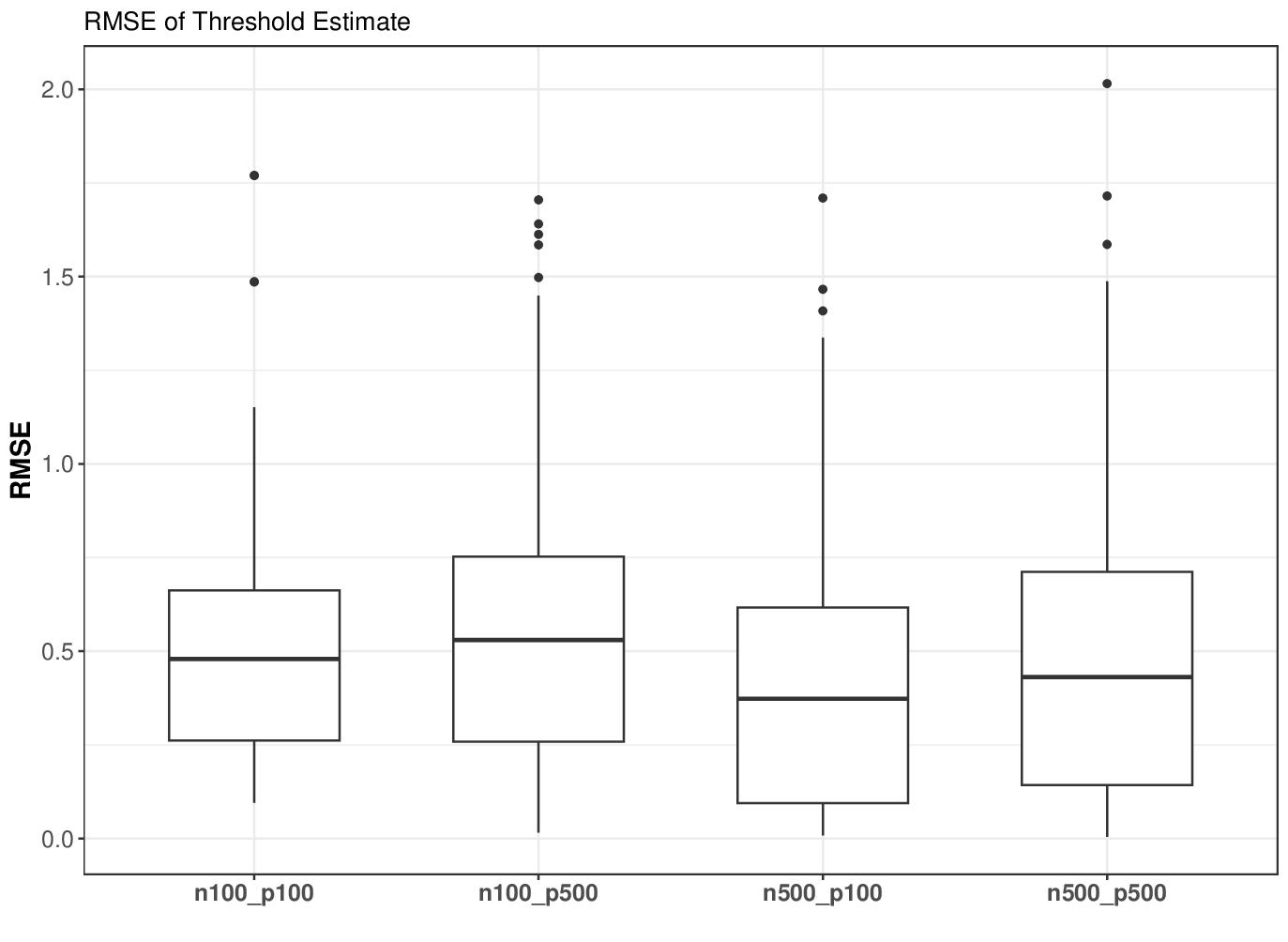}
    \caption{RMSE of the estimated thresholds using Algorithm 1 across settings with $n = 100, 500$ and $p = 100, 500$.}
    \label{fig:threshold}
\end{figure}

To evaluate the accuracy of the thresholds $\widehat{T}$,  we simulated mixed data across DAG settings with $n = 100, 500$ and $p = 100, 500$. The root mean squared error between final threshold estimates $\widehat{T}_j$ and true thresholds $T_j = \{-1,1\}$ was computed as
\begin{align}
    \text{RMSE}(\widehat{T}) = \left( \frac{1}{|T|}\sum_{j,c} (\widehat{\tau}_{j,c} - \tau_{j,c})^2 \right)^{1/2},
\end{align}
where $|T|$ denotes the number of estimated thresholds, $\tau_{j,1} = -1$ and $\tau_{j,2} = 1$. Figure~\ref{fig:threshold} shows that, across settings with $n = 100,500$ and $p = 100,500$, the average $\text{RMSE}(T)$ does not vary significantly. 

\begin{figure}[htbp]
    \centering
    \includegraphics[width=0.75\linewidth]{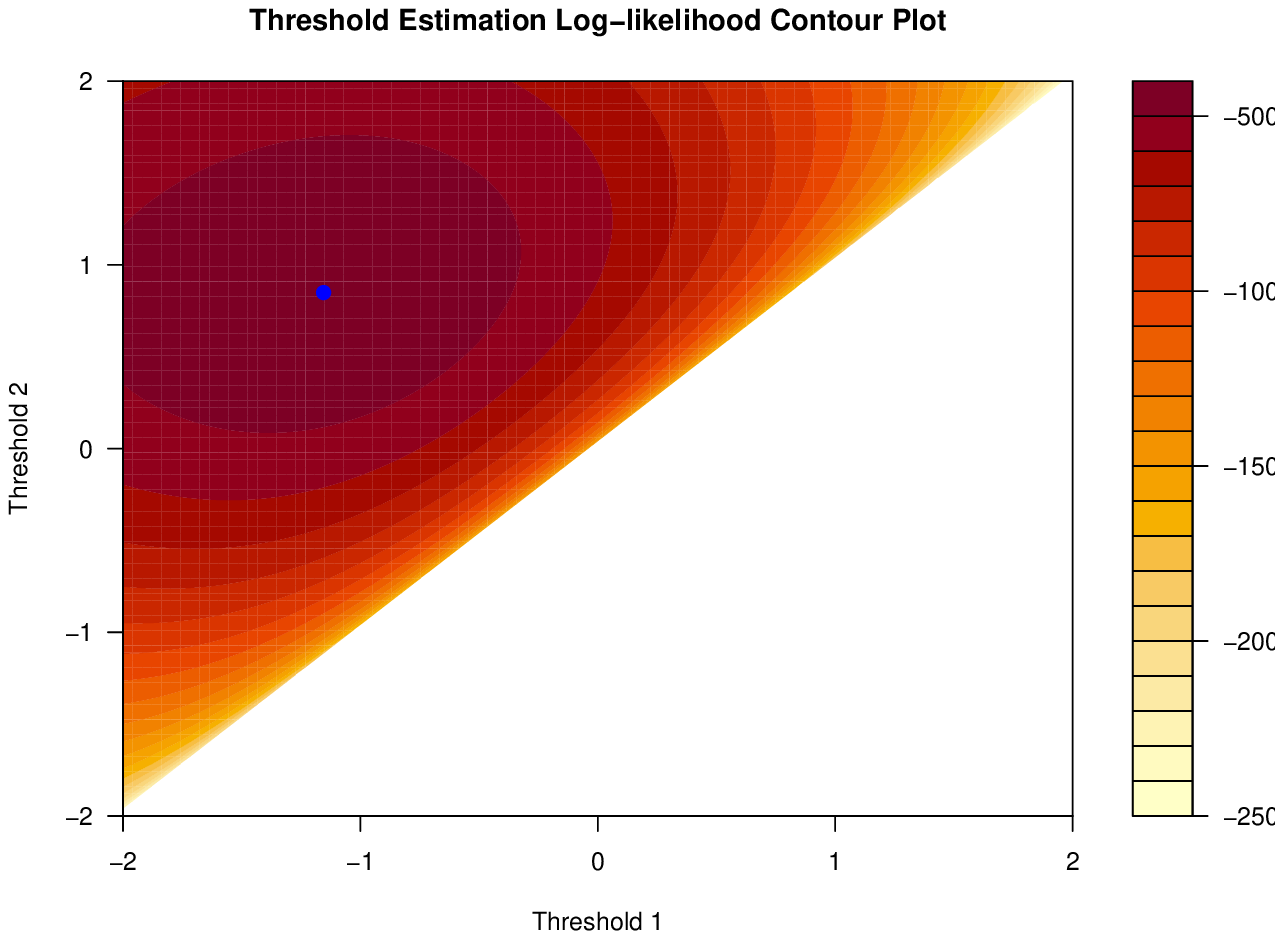}
    \caption{Log-likelihood contour plot of the thresholds for one discrete variable with $n = 500$ and $p = 500$. The blue dot indicates the maximizer, which is close to the true value $(-1,1)$.}
    \label{fig:contour}
\end{figure}

The log-likelihood contour plot in Figure~\ref{fig:contour} shows a relatively flat likelihood surface, suggesting substantial variance in the threshold estimates. While the estimated thresholds are not always close to the true values, the accuracy of the covariance estimates and subsequent causal graph estimation indicates that the estimated thresholds were sufficient to support the rest of the causal graph learning procedure.

\section{Additional results for the GRN application}

This section provides additional results of the scRNA-seq data application presented in Section~6 of the main paper. 

\begin{figure}[htbp]
    \centering
    \includegraphics[width=0.75\linewidth]{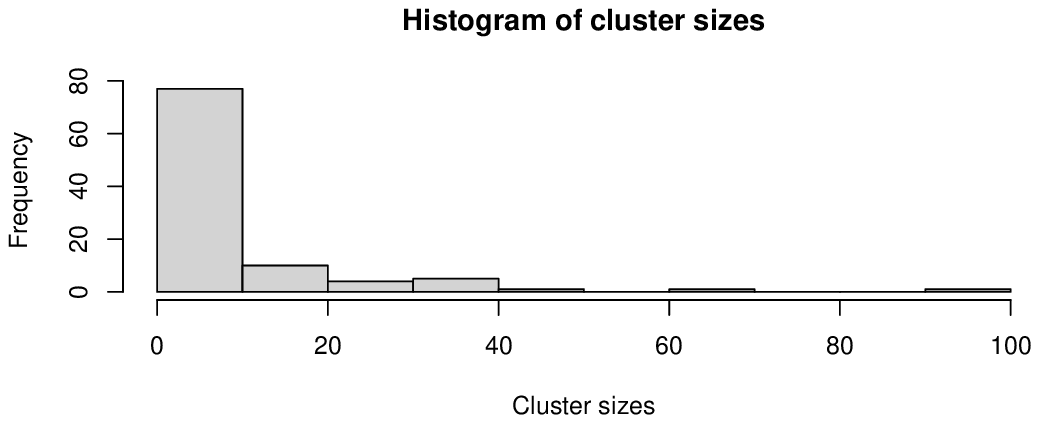}
    \caption{Histogram of cluster sizes using hierarchical clustering on all $n = 859$ cells utilizing 2,000 background genes.}
    \label{fig:hist_cluster_sizes}
\end{figure}

We first show the distribution of cluster sizes in Figure~\ref{fig:hist_cluster_sizes} obtained via hierarchical clustering of $n = 859$ cells into 100 clusters. 

\begin{figure}[htbp]
    \centering   
    \includegraphics[width=0.75\linewidth, trim = 0 0 0 0.5in, clip]{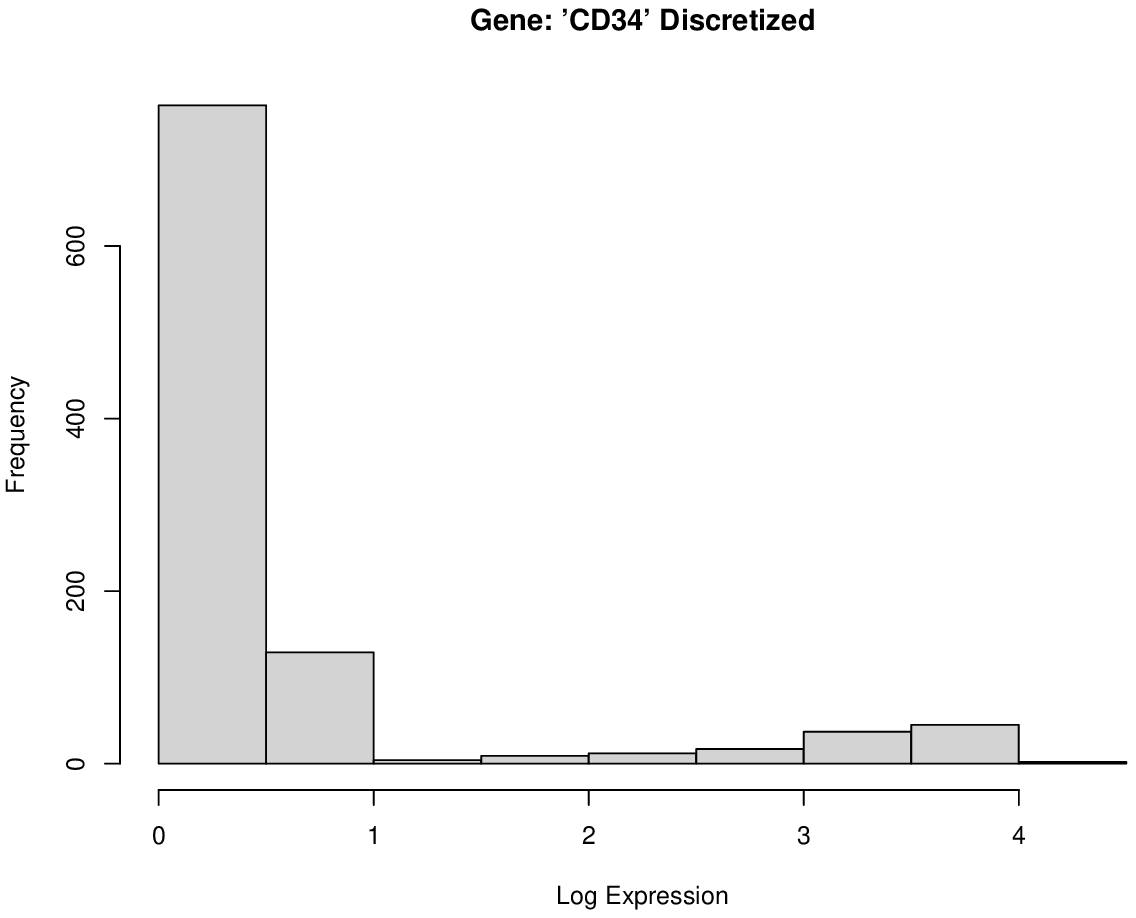}
    \caption{Gene expression pattern of  CD34 that is discretized by our criteria.}
    \label{fig:cd34_discrete}
\end{figure}

The criteria used to determine whether or not individual genes are to be discretized is explained in Section~6.2 of the main paper. Figure~\ref{fig:cd34_discrete} shows the expression profile of CD34, which is discretized into two levels.

\begin{figure}[htbp]
    \centering
    \includegraphics[width=0.75\linewidth,trim = 0 0 0 0.5in, clip]{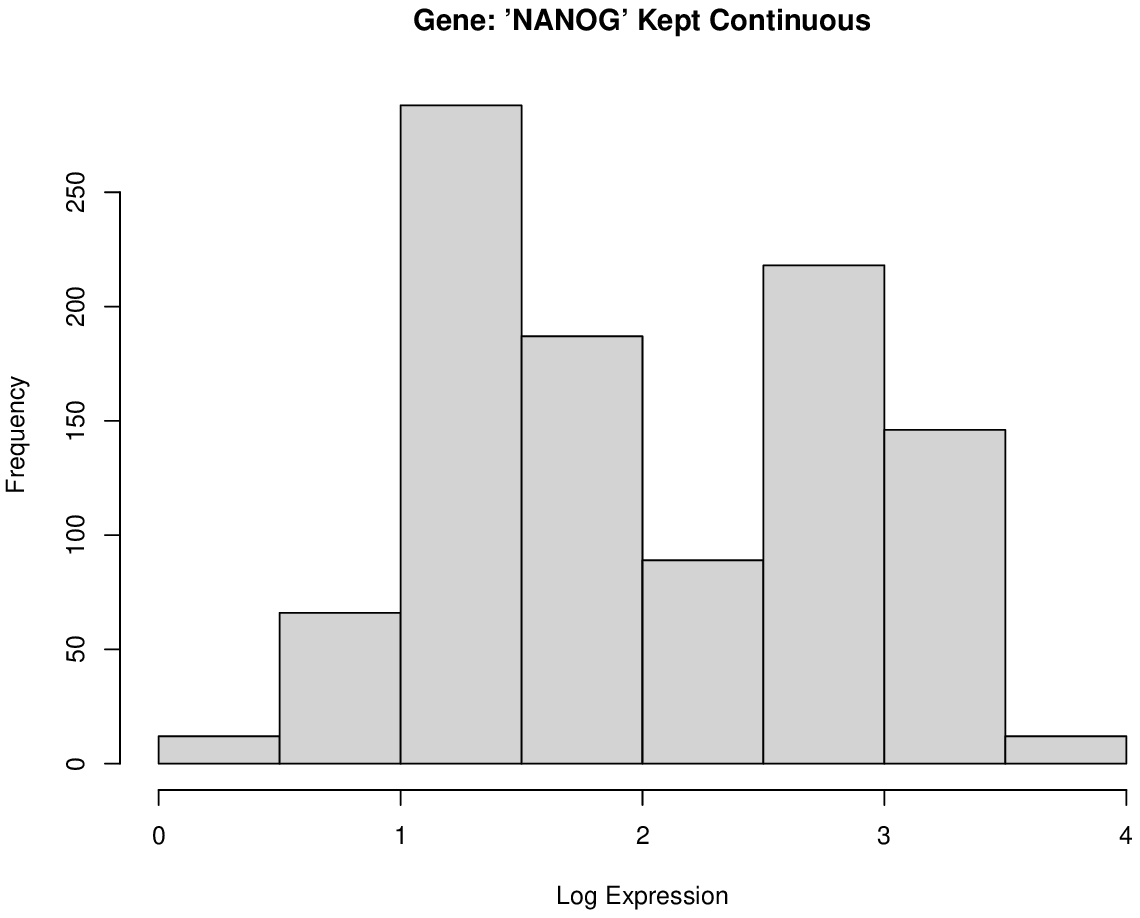}
    \caption{Gene expression pattern of NANOG that is kept continuous.}
    \label{fig:nanog_continuous}
\end{figure}

In contrast, other genes exhibiting more smooth uni-modal or mixture Gaussian expression patterns are retained as continuous variables. Figure~\ref{fig:nanog_continuous} shows the expression pattern of NANOG as an example.
\newpage

Finally, we report in Table~\ref{table:lit_genes} the full set of predicted edges inferred from the scRNA-seq data via bootstrap resampling. This table complements the summarized network results shown in the main paper and provides all inferred edges with confidence score $> 50\%$.
\begin{table}[htbp]
\centering
\caption{Full list of predicted edges from the gene regulatory network using bootstrap resampling}
\label{parset}
\begin{tabular}{lll}
\hline
$X_i - X_j$ & Confidence \% & Reference \\
\hline
POU5F1 $-$ ZFP42  & 100.0 & \citet{wang2006protein}\\
NANOG $-$ POU5F1 & 99.0  & \citet{mitsui2003homeoprotein}\\
LEFTY1 $-$ NODAL  & 98.5 & \citet{juan2001roles}\\
PECAM1 $-$ POU5F1 & 98.5  \\
PECAM1 $-$ ZFP42  & 97.5 \\
DNMT3B $-$ ZFP42  & 96.5 \\
DNMT3B $-$ LECT1  & 91.0  \\
CER1   $-$ EOMES  & 84.5 \\
DNMT3B $-$ POU5F1 & 84.5 & \citet{liu2025dna}\\
LECT1  $-$ ZFP42  & 79.5 \\
LEFTY2 $-$ PMAIP1 & 77.0 \\
GATA6  $-$ SOX2   & 70.5 \\
IGFBP3 $-$ LEFTY1 & 59.5 \\
GABRP  $-$ PAX6   & 57.0 \\
GDF3   $-$ PMAIP1 & 57.0 \\
MT1X   $-$ PMAIP1 & 50.5 \\
NODAL  $\rightarrow$ PMAIP1 & 73.0 \\
PAX3   $\rightarrow$ MAP2   & 72.5 \\
PAX6   $\rightarrow$ PAX3   & 71.3 \\
HAPLN1 $\rightarrow$ GNG11  & 66.5 \\
HNF4A  $\rightarrow$ FOXQ1  & 66.9 \\
GDF3   $\rightarrow$ DNMT3B & 66.3 \\
DLK1   $\rightarrow$ PAX3   & 64.5  \\
TNFSF10 $\rightarrow$ PECAM1 & 62.0  \\
PRDM14 $\rightarrow$ NANOG  & 61.2 & \citet{ma2011sequence}\\
MAPK10 $\rightarrow$ PAX3   & 61.0 \\
CD34   $\rightarrow$ PECAM1 & 61.0  \\
HAND1  $\rightarrow$ SOX2   & 60.0  \\
SOX9   $\rightarrow$ MAP2   & 58.3  \\
GATA2 $\rightarrow$ HAPLN1 & 57.5  \\
PAX3   $\rightarrow$ ZFHX4  & 57.3 \\
MT1X   $\rightarrow$ DNMT3B & 56.0 \\
ZSCAN10 $\rightarrow$ PMAIP1 & 56.0  \\
PMAIP1 $\rightarrow$ POU5F1 & 55.3 \\
PMAIP1 $\rightarrow$ LECT1  & 54.5 \\
LHX1   $\rightarrow$ SOX17  & 53.0 \\
GATA4  $\rightarrow$ FOXQ1  & 51.0 \\
\hline
\end{tabular}
\label{table:lit_genes}
\end{table}
\bibliographystyle{imsart-nameyear} 
\bibliography{ref}